\documentclass[fleqn,usenatbib,useAMS]{mnras}

\usepackage[T1]{fontenc}

\DeclareRobustCommand{\VAN}[3]{#2}
\let\VANthebibliography\thebibliography
\def\thebibliography{\DeclareRobustCommand{\VAN}[3]{##3}\VANthebibliography}

\usepackage{graphicx}	
\usepackage{amsmath}	
\usepackage{amssymb}	
\usepackage[utf8]{inputenc}

\usepackage{multirow}

\newcommand{\ujyperbeam}{$\mu$Jy\,beam$^{-1}$}

\newcommand{\kms}{km s$^{-1}$}

\usepackage[dvipsnames]{xcolor}

\title[IRAS 16293B hosts a hot and gravitationally unstable disk]{The young protostellar disk in IRAS16293-2422 B is hot and shows signatures of gravitational instability}

\author[Joaquin Zamponi et al.]{
Joaquin Zamponi,$^{1}$\thanks{E-mail: jzamponi@mpe.mpg.de}
Mar\'ia Jos\'e Maureira,$^{1}$
Bo Zhao,$^{1}$
Hauyu Baobab Liu,$^{2}$
John D. Ilee,$^{3}$
\newauthor
Duncan Forgan,$^{4}$ 
and Paola Caselli$^{1}$
\\
$^{1}$Max-Planck-Institut f\"ur Extraterrestrische Physik (MPE), D85748 Garching, Germany\\
$^{2}$Institute of Astronomy and Astrophysics, Academia Sinica, 11F of Astronomy-Mathematics Building, AS/NTU No.1, Sec. 4, \\
$^{\,\,}$Roosevelt Rd, Taipei 10617, Taiwan, ROC\\
$^{3}$School of Physics and Astronomy, University of Leeds, Leeds LS2 9JT, UK\\
$^{4}$Centre for Exoplanet Science, SUPA, School of Physics \& Astronomy, University of St Andrews, St Andrews KY16 9SS, UK
}

\date{Accepted XXX. Received YYY; in original form ZZZ}

\pubyear{2021}

\usepackage{newtxtext,newtxmath}

\begin{document}
\label{firstpage}
\pagerange{\pageref{firstpage}--\pageref{lastpage}}
\maketitle

\begin{abstract}
Deeply embedded protostars are actively fed from their surrounding envelopes through their protostellar disk. The physical structure of such early disks might be different from that of more evolved sources due to the active accretion. We present 1.3 and 3\,mm ALMA continuum observations at resolutions of 6.5\,au and 12\,au respectively, towards the Class 0 source IRAS 16293-2422 B. 
The resolved brightness temperatures appear remarkably high, with $T_{\rm b} >$ 100\,K within $\sim$30\,au and $T_{\rm b}$ peak over 400\,K at 3\,mm. 
Both wavelengths show a lopsided emission with a spectral index reaching values less than 2 in the central $\sim$ 20\,au region. We compare these observations with a series of radiative transfer calculations and synthetic observations of magnetohydrodynamic and radiation hydrodynamic protostellar disk models formed after the collapse of a dense core. 
Based on our results, we argue that the gas kinematics within the disk may play a more significant role in heating the disk than the protostellar radiation. 
In particular, our radiation hydrodynamic simulation of disk formation, including heating sources associated with gravitational instabilities, is able to generate the temperatures necessary to explain the high fluxes observed in IRAS 16293B.  
Besides, the low spectral index values are naturally reproduced by the high optical depth and high inner temperatures of the protostellar disk models. The high temperatures in IRAS 16293B imply that volatile species are mostly in the gas phase, suggesting that a self-gravitating disk could be at the origin of a hot corino.
\end{abstract}




\section{Introduction}
\label{sec:introduction}
In the earliest stages of star formation, the Class 0 stage \citep{Andre2000}, the protostellar envelope still contains a significant fraction of the total mass of the system. 
The protostellar disk is continuously fed by the surrounding envelope and it may become gravitationally unstable depending on environmental conditions at the start of the pre-stellar core collapse \citep[e.g.][]{Zhao2018}. 
Large streamers of molecular material from the outer envelope and surrounding cloud can also contribute to the disk mass growth \citep{Dullemond2019,Akiyama2019,Pineda2020,Kueffmeier2020}. 
Gravitationally unstable disks can fragment, enabling the formation of giant planets \citep{Boss2009,Vorobyov2010,Machida2011a} and develop spiral arms where gas compression and shocks locally heat the gas and dust to values well above those predicted by irradiated viscously evolving disks \citep[e.g.][]{BoleyAndDurisen2008,Dong2016}. 
The chemical composition of the disk is heavily affected by gravitational instabilities (GI) \citep[e.g.][]{Ilee2011, Ilee2017}. 
Moreover, their relatively large masses and surface mass densities of gravitationally unstable disks imply large opacities of their dust emission at millimeter and sub-millimeter wavelengths, thus hindering total mass measurements \citep[e.g.][]{Evans2017,Galvan-Madrid2018,Li2017} and making molecular line observations difficult to probe their full structure \citep{Evans2019}. 

So far, there is no clear evidence of a gravitational unstable disk among Class 0 sources, with the possible exception of the disk around the triple protostar system L1448 IRS3B \citep{Tobin2016}, but this disk is highly perturbed by the presence of multiple sources which can induce tidal forces and instabilities mimicking original GIs. To make progress in this field and unveil a gravitationally unstable disk at the earliest stages of star formation, one needs high sensitivity and high angular resolution observations of a bright Class 0 source and state-of-the art numerical simulations of disk formation which can then be compared in detail with observations. 

Located in the star-forming region $\rho$-Ophiuchi, inside the dark cloud L1689N and at a distance of 141~pc \citep{Dzib2018}, IRAS16293-2422 is a well studied Young Stellar Object (YSO) classified as a Class 0 source with less than $10^4$~yr \citep{Andre1993}, and represents one of the very early stages of low-mass star formation.  
It was the first source identified as a hot corino \citep{Blake1994, vanDischoek1995} based on the detection of Complex Organic Molecules (COMs) in the source, which was later supported by follow up studies (\citealt{Ceccarelli1998,Ceccarelli2000,Schoeier2002, Crimier2010, Pineda2012, Jorgensen2011, Jorgensen2016, Oya2016, Jacobsen2018, vanDerWiel2019}).  
Higher resolution observations revealed that IRAS16293-2422 is in fact a triple system, composed of sources A1 and A2, separated by 54~au from each other \citep{Maureira2020} and source B, 738~au (5"; \citealt{Wootten1989}) away from source A. 
Due to this larger separation, tidal truncation between the three protostars is discarded and therefore source B is considered to have evolved as an isolated source \citep{Rodriguez2005}.  
It was initially proposed to be either an evolved T Tauri star \citep{Stark2004,Takakuwa2007} or a very young object \citep{Chandler2005}, however, \citet{Chandler2005} suggested that source B has large scale infalls based on SO line emission. 
\citet{Pineda2012} confirmed the infall of an inner envelope, with mass accretion rates of $4.5\times10^{-5}$~M$_{\odot}{\rm yr}^{-1}$, based on ALMA detections of inverse P-Cygni profiles in CH$_3$OCHO-E, CH$_3$OCHO-E-A and H$_2$CCO, ruling out the possibility of it being a T Tauri star. 
The interpretations of infall from these profiles was also suggested by \citet{Jorgensen2012} and \citet{Zapata2013}. 
Unlike the A1 and A2 protostars, source B has not shown clear signs of outflow launching, explained by the lack of free-free emission at low frequencies \citep{Chandler2005, Rodriguez2005, Loinard2007, Rao2009,Liu2018b, Hernandez-Gomez2019b} and also based on molecular lines \citep{Loinard2002,vanDerWiel2019}.

The possibility of source B being gravitationally unstable in the outer part of its disk was also proposed by \citet{Rodriguez2005} and discussed by \citet{KratterAndLodato2016}.  
The disk around source B must be highly dense and massive ($\gtrsim$0.2~M$_{\odot}$; \citealt{Rao2009,Pineda2012}) 
to produce gravitational instabilities. 
After comparing high sensitivity interferometric data with MHD and RHD star formation simulations, we demonstrate here that indeed the physical structure (in particular the temperature) of the IRAS16293-2422 B disk resembles that of a self-gravitating disk.  

This paper is organized as follows: in section \ref{sec:ALMA_observations} we provide details on the data observation and reduction, followed by our results from the ALMA observations, in section \ref{sec:Protostellar_disk_simulations} we describe the numerical simulations used to compare against our ALMA observations, in section \ref{sec:Radiative_transfer} we outline the post-processing scheme done with radiative transfer calculations, in section \ref{sec:results} we compare our synthetic observations to the real data, then we discuss about possible interpretations for the features observed in section \ref{sec:discussion} and finally, we present our conclusion in section \ref{sec:Conclusions}.

\section{Observations}
\label{sec:ALMA_observations}

\subsection{Data and Imaging}
\label{sec:data_calibration}
Observations of IRAS 1629-2422 were taken in band 3 and band 6 with ALMA (project IDs: 2017.1.01247.S and 2016.1.00457.S, respectively). The band 3 observations were taken on October 8 and 12, 2017 in the most extended Cycle 5 configuration with a baseline range of 41.4 m–16.2 km and a maximum recoverable scale of $\sim$0.5" (or 70 au). The single spectral window for the band 3 continuum used in this work has a bandwidth of 2 GHz divided into 128 channels and centered at 99.988 GHz. The band 6 observations were taken during 2017 August 21. The baselines range between 21 and 3697 m, with a maximum recoverable scale of $\sim$0.6" (or 85 au). The spectral setup consisted of four spectral windows centered at a frequency of 240.3 GHz, 240.5 GHz, 224.7 GHz, 222.9 GHz, with a channel width of 61.0 kHz, 15.3 kHz, 30.5 kHz, 488.3 kHz and bandwidth of 0.23 GHz , 0.06 GHz,  0.11 GHz, 0.94 GHz, respectively. To create the continuum image we carefully check each spectral window and flag the lines. The resulting bandwidth for the band 6 continuum is 0.12 GHz with most of the channels coming from the spectral window centered at 222.9 GHz. Both, the band 3 and band 6 continuum data has been previously published in \cite{Maureira2020} and \cite{Oya2020}, respectively.

For band 3, the procedure for the calibration of the continuum data (including phase and amplitude self-calibration) as well as the imaging are detailed in \cite{Maureira2020}. In summary, when imaging the continuum we iteratively performed phase-only self-calibration with a minimum solution interval of 9 seconds. Afterwards we performed two amplitude self-calibration iterations, with a minimum solution interval of 60 seconds. The final continuum dataset after phase+amplitude self-calibration was imaged using the \texttt{tclean} task from the Common Astronomy Software Applications (CASA; v5.6.2) with the multiscale deconvolver \citep{Cornwell2008MultiscaleClean,Rau2011MultiscaleClean} and a robust parameter of 0.5. The beam size, beam position angle (P.A.) and noise of the continuum image are 0.048"$\times$0.046" (6.5 au), 79.3$^{\circ}$ and 15 $\mu$Jy\,beam$^{-1}$, respectively. This image was used when analyzing the band 3 data alone, while a new additional map was done for producing a spectral index map (see below). For band 6, we perfomed iteratively phase-only self-calibration with a minimum solution interval of \texttt{solint='int'} or 4 seconds. Afterwards we performed one amplitude self-calibration, with \texttt{solint='inf'} (two solutions, one for each track). For the process of self-calibration of the final image we selected visibilities with a minimum baseline of 120 klambda, in order to avoid missing flux artifacts, but also because shorter baselines are not covered by the band 3 observations with which we want to compare for our analysis. 

The continuum image for band 6 and an additional continuum image for band 3 with matching \textit{uv}-range were created using \texttt{tclean}. This additional image of the band 3 data with matching uv-range was done with the goal of producing a spectral index map. We use multiscale, a robust parameter of 0 and a \textit{uv}-range parameter of 120-2670 $k\lambda$ (overlapping baselines) for both datasets. A reference frequency of 223 GHz and 100 GHz was set for the band 6 and band 3 observations, respectively. The resultant beam size, P.A. and rms of the band 3 observations correspond to 0.062"$\times$0.050", 41.7$^{\circ}$ and 17 \ujyperbeam, respectively. Similarly, for the band 6 observations these values are 0.114"$\times$0.069", -88.2$^{\circ}$ and 104 \ujyperbeam, respectively. This study focuses on source B, corresponding to the single northern source of the IRAS 16293-2422 triple system. The corresponding maps for the southern pair (A1 and A2), as well as further details of the imaging in both bands, will be presented in Maureira et al. (in prep).

\subsection{Results}
\label{sec:observational_results}

\subsubsection{Continuum observations at 1.3 and 3 mm}
\label{sec:continuum_observations}

The continuum observations obtained with ALMA toward source B are shown in Fig.~\ref{fig:IRAS16293B_observations} at $\lambda=1.3$ and 3~mm.  The structure seen at 3~mm has a radius of $\sim$46~au, measured along the major axis of a contour at 5$\,\sigma$. This size is consistent with the reported typical sizes of embedded disks \citep{Segura-Cox2018,Maury2019, Tobin2020a} and also consistent with previous observations of this source \citep{Rodriguez2005, Oya2018}. 

Interestingly, both wavelengths show the emission peak clearly shifted to the West of the center of the overall structure. 
At a first glance, this shift could imply a possible inclination of the disk or the presence of a real asymmetry in density and/or temperature. The aspect ratio of the disk is $\sim$0.95 (derived from the east-west over north-south extensions of source B outlined by a contour at 5$\,\sigma$). 
The ratio was measured in the the optically thinner 3~mm image which, under the assumption of a circular disk, suggests a moderate inclination of $\sim$18$^{\circ}$. 
If the disk were instead highly inclined, it would be expected that the peak would be more centered when seen at longer wavelengths. However, hints of a shift were also seen in high-resolution VLA continuum observations at 42~GHz reported by \citet{Rodriguez2005} and \citet{Hernandez-Gomez2019b}, unlike what is seen in the case of the highly inclined disk in HH212 \citep{Lin2021}. 
Furthermore, the line emission from complex organic molecules from the ALMA-PILS survey also appears to show a peak shifted to the west \citep{Calcutt2018a,Calcutt2018b,Manigand2020,Manigand2021}. Likewise, the source does not show a significant gradient indicative of rotation in previous ALMA line observations from 100 au down to 70 au scales. Instead, the kinematics is consistent with mostly infall motions \citep{Pineda2012,Zapata2013,Oya2018} in a system with a face-on disk. All the above evidence suggests that there might be a real asymmetry present in the temperature and/or density structure of source B, instead of being a product of optical depth and inclination. 

The brightness temperatures and rms uncertainties in the position of the peak are $T_{\rm b}^{\rm 1.3mm}=287\pm0.3$~K and $T_{\rm b}^{\rm 3mm}=470\pm1$~K, with the 3mm one remaining higher than the 1.3mm even after smoothing the 3mm observations to match the beam of the 1.3mm observations ($T_{\rm b,\, smooth}^{\rm 3mm}=364\pm0.4$~K). This same trend, of higher brightness temperature for longer wavelengths, is also consistent with recent VLA observations at 41~GHz (7~mm) and 33~GHz (9~mm) in \citet{Hernandez-Gomez2019b}, which had a comparable resolution to our observations. They measured a peak brightness of $T_{\rm b}^{\rm 7mm}=870$~K and $T_{\rm b}^{\rm 9mm}=700$~K. Additionally, they reported ALMA observations at 700~GHz that have a brightness temperature of  $T_{\rm b}^{\rm 1mm}=185$~K. This trend is consistent with self-obscuration of inner hot material due to high optical depth, a scenario that we investigate in Section \ref{sec:simulated_observations_grav-unstable_model} and discuss in Section \ref{sec:are_early_disks_hotter_in_the_midplane}.  

\subsubsection{Spectral index}
\label{sec:spectral_index_ALMA}

\begin{figure*}
    \centering
    \includegraphics[width=18cm, trim=1cm 0 1cm 0cm, clip]{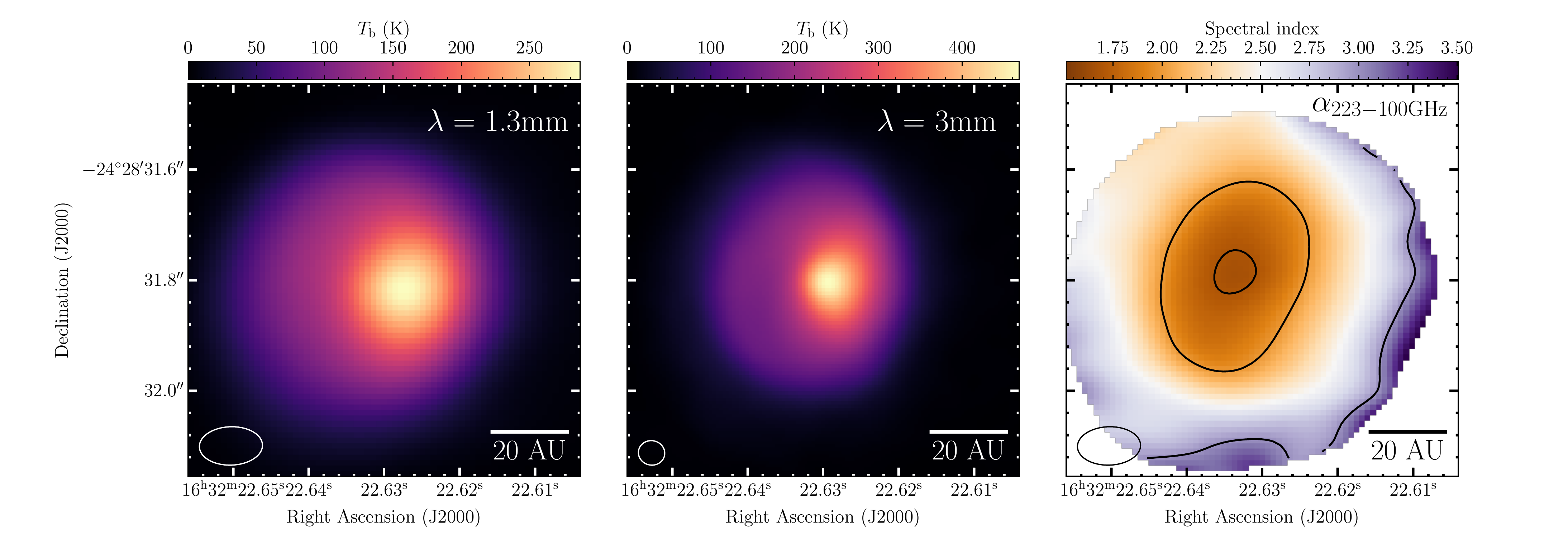}
    \caption{ALMA observations of IRAS16293-2422B. The left and central panels show the brightness temperatures observed at 1.3\,mm (100\,GHz; band 6) and 3\,mm (223\,GHz; band 3), respectively. The observation at 3\,mm report brighter emission than at 1.3\,mm, with a peak value of $T_{\rm b}\sim470\,$K and $T_{\rm b}\sim290\,$K, respectively. Both images indicate the presence of a resolved brightness asymmetry in the source. The rightmost panel shows the spectral index calculated between the two wavelengths (as described in Appendix~\ref{appendix:spectral_index}), for which the 3\,mm image was smoothed to match the 1.3\,mm beam. Contours for the spectral index map are shown at 1.7, 2 and 3, inside-out. The synthesized beams are 0.114"$\times$0.069" and 0.048"$\times$0.046" at 1.3 and 3~mm (see bottom left corners).}
    \label{fig:IRAS16293B_observations}
\end{figure*}

We derived the spectral index $\alpha$ between our ALMA observations at 1.3 and 3~mm using equation \ref{eq:spectral_index}. For this, we use maps imaged using the same \textit{uv}-range (Section~\ref{sec:data_calibration}) and smoothing afterwards to match beams. The results are shown on the rightmost panel of Fig.~\ref{fig:IRAS16293B_observations}. Black contours are placed at $\alpha$=1.7, 2 and 3, inside-out. A clear decrease of spectral index is observed towards the center, ranging from 3 to values as low as 1.7. The middle contour, at $\alpha$=2, shows an elongated shape. A similarly low $\alpha$ was measured in the center by \citet{Loinard2007} and it was suggested to be due to a modest thermal jet. However, \citet{Hernandez-Gomez2019b} compiled integrated fluxes using SMA, VLA and ALMA observations ranging from 3 to 700~GHz and found no evidence of free-free emission. The SED was well fitted with a single spectral index of  $\alpha=2.28\pm0.02$, in agreement with optically thick thermal dust emission at all wavelengths. 

Spectral indices close to and lower than 2 have also been reported toward other sources, such as those in the PROSAC survey of Class 0/I protostars \citep{Jorgensen2007,Jorgensen2009} and also toward the Class I source WL12 \citep{Miotello2014}. More recently, \citet{Lin2021} presented a map of the spectral index between ALMA bands 3, 6 and 7 towards the embedded edge-on disk around the Class 0 protostar HH212. The spectral index reaches values as low as $\sim$1.5 within the central $\sim$50 ($\alpha_{\rm band\,3,6}$) and 80\,au ($\alpha_{\rm band\,6,7}$). The observations were well reproduced by considering a temperature gradient increasing toward the inner regions of the disk, which under optically thick conditions, results in self-obscuration of the inner hot regions and thus in low spectral indexes, as also suggested by \cite{Li2017} and \cite{Galvan-Madrid2018}. In section~\ref{sec:are_early_disks_hotter_in_the_midplane}, we discuss further the origin of the low spectral index in source B, based on comparison with protostellar disk models formed from the collapse of a dense core.

\section{Protostellar disk simulations}
\label{sec:Protostellar_disk_simulations}

\subsection{Simulations of disk formation and evolution}
\label{sec:Simulations_of_disk_formation}

\subsubsection{non-ideal MHD disk model}
\label{sec:non-ideal_MHD_model}
In this work we have post-processed 3D non-ideal magnetohydrodynamic (MHD) simulations of protostellar disk formation presented in~\citet{Zhao2018}, using the code ZeusTW~\citep{Krasnopolsky2010}. 
They employed a spherical grid that is non-uniform along the radial axis, to provide high resolution in the inner regions of the grid. 
The grid boundary has a 2\,au radius and a 6684\,au outer radius.  
All gas that flows into this 2\,au central cell is considered to be accreted by the protostar.
No additional sink particles were included in the simulations.  
These simulations start from the cloud core scale and evolve up to $\sim$20\,kyr after the formation of a rotationally supported disk, during which ambipolar diffusion plays the dominant role.
We have selected the model \textit{2.4Slw-trMRN} from Table 1 in~\citet{Zhao2018}.
This core is initialized as an isolated isothermal sphere of uniform density ($\rho_0=4.77\times10^{-19}$\,g\,cm$^{-3}$), with core radius $R_{\rm c}=6684$\,au, core mass $M_{\rm c}=1$\,M$_{\odot}$ and a constant temperature of 10\,K.
Assuming a mean molecular mass of $\mu=2.36$ amu, the gas number density for molecular hydrogen correspond to $n(H_2)=1.2\times10^5$~cm$^{-3}$. 
The core initially rotates as a solid body with angular speed $\omega_0=10^{-13}$~s$^{-1}$, which leads to a rotational to gravitational energy ratio of $\beta_{\rm rot}=0.025$.
The core is magnetized and the initial magnetic field is uniformly distributed, along the rotation axis, having a constant strength of $B_0=4.25\times10^{-5}\,$G, which corresponds to a dimensionless mass-to-flux ratio of $\lambda=2.4$.

\begin{figure*}
     \centering
     \includegraphics[width=16cm, trim=0cm 5.95cm 0cm 3.35cm, clip]{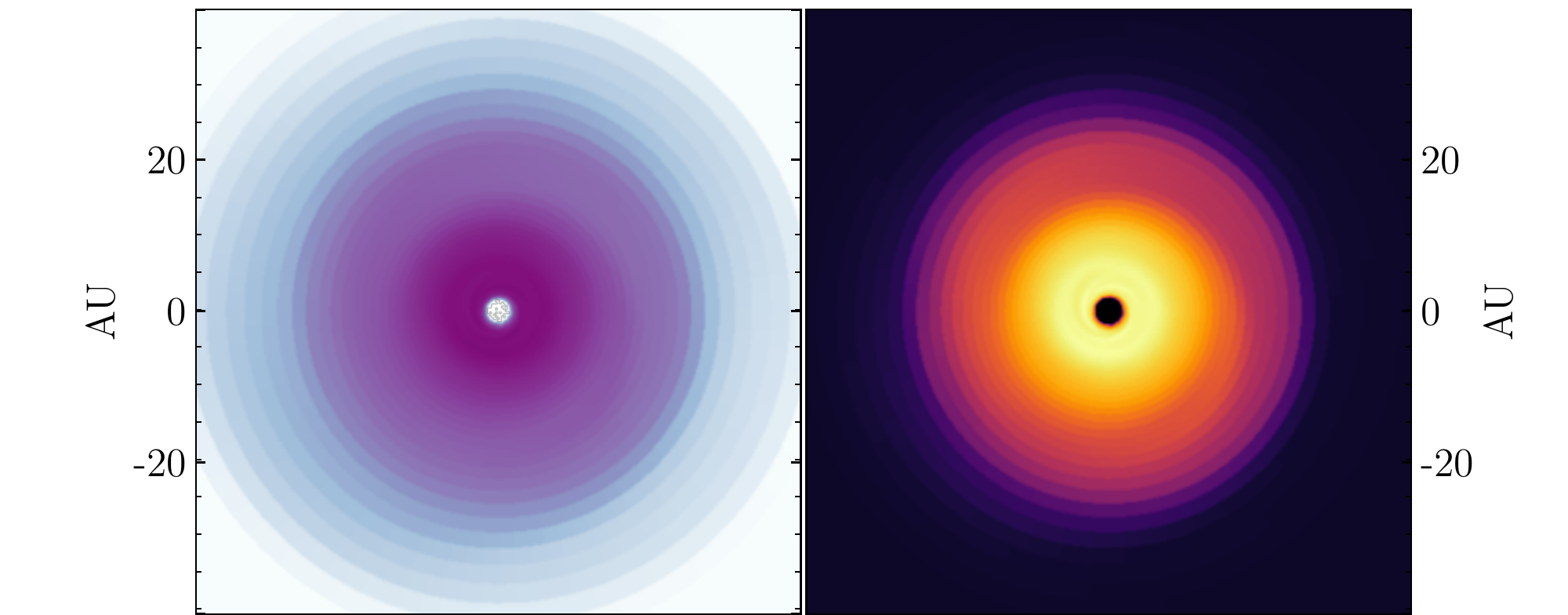}
     \includegraphics[width=16cm, trim=0cm 0.5cm 0cm 12.95cm, clip]{img/disk_model_bo.pdf}\!\!
     \caption{Face- and edge-on slices of the gas density (left column) and gas temperature (right column) distributions for the MHD disk model from \citet{Zhao2018}, as described in section \ref{sec:non-ideal_MHD_model}.}
     \label{fig:bo_model}
 \end{figure*}

The selected simulation model shows a relatively compact disk of $\sim$50~au in diameter, without obvious spiral structures, due to the strong magnetic field and slow rotation of the initial core. 
We selected the timestep at which the disk mass was the highest (0.09 M$_{\odot}$). The disk is only marginally gravitationally unstable (see Fig. \ref{fig:toomre_q}) and relatively smooth, although at earlier times developing transient spiral arms.
This snapshot represents a state $\sim$144~kyr after the beginning of the core collapse and $\sim$22.34~kyr after the formation of a first core \citep{Larson1969}. 
Face-on and edge-on slices of the gas density and gas temperature distributions of the MHD disk model are shown in Fig.~\ref{fig:bo_model}. 

Starting from the inner boundary, the radial density distribution on the midplane goes as follows: the maximum is reached at the inner boundary region of 2-3 au with a density of $4.5\times 10^{-11}$\,g\,cm$^{-3}$ ($n_{\rm gas} = 1.14\times10^{13}$ cm$^{-3}$). The density remains roughly constant up to 7~au and then falls as a power-law of $r^{-2.8}$ until the disk's outer edge at 25~au. The overall density distribution is rather smooth across the entire disk with no signs of fragmentation. 
The vertical density distribution shown in the edge-on view of Fig.~\ref{fig:bo_model}, indicates the presence of a midplane much denser than the outer layers, especially within the inner 7\,au and a scale height of $\lesssim5\,$au. 

The initial stages of the collapse proceeded isothermally at 10\,K and then followed an adiabatic equation of state (EOS; see Appendix A in \citealt{Zhao2018}) after a density threshold of 10$^{-13}$\,g\,cm$^{-3}$ was reached. 
The gas temperature is then obtained from a barotropic EOS with an adiabatic index of 5/3 (up to $10^{-11}\,$g\,cm$^{-3}$) and therefore follows always the density distribution. 
The edge-on view of the temperature (see lower-right panel in Fig.\ref{fig:bo_model}) shows a gradient from the outer layers toward the midplane. 
In this warm midplane, the hottest regions are at $T$$\sim$$190$\,K and lie within the densest 7\,au radius. 

The edge-on view of the disk also show bipolar outflows and accreting flows from the parent dense core. 
 
 \begin{figure*}
     \centering
     \includegraphics[width=16cm, trim=0cm 5.95cm 0cm 3.35cm, clip]{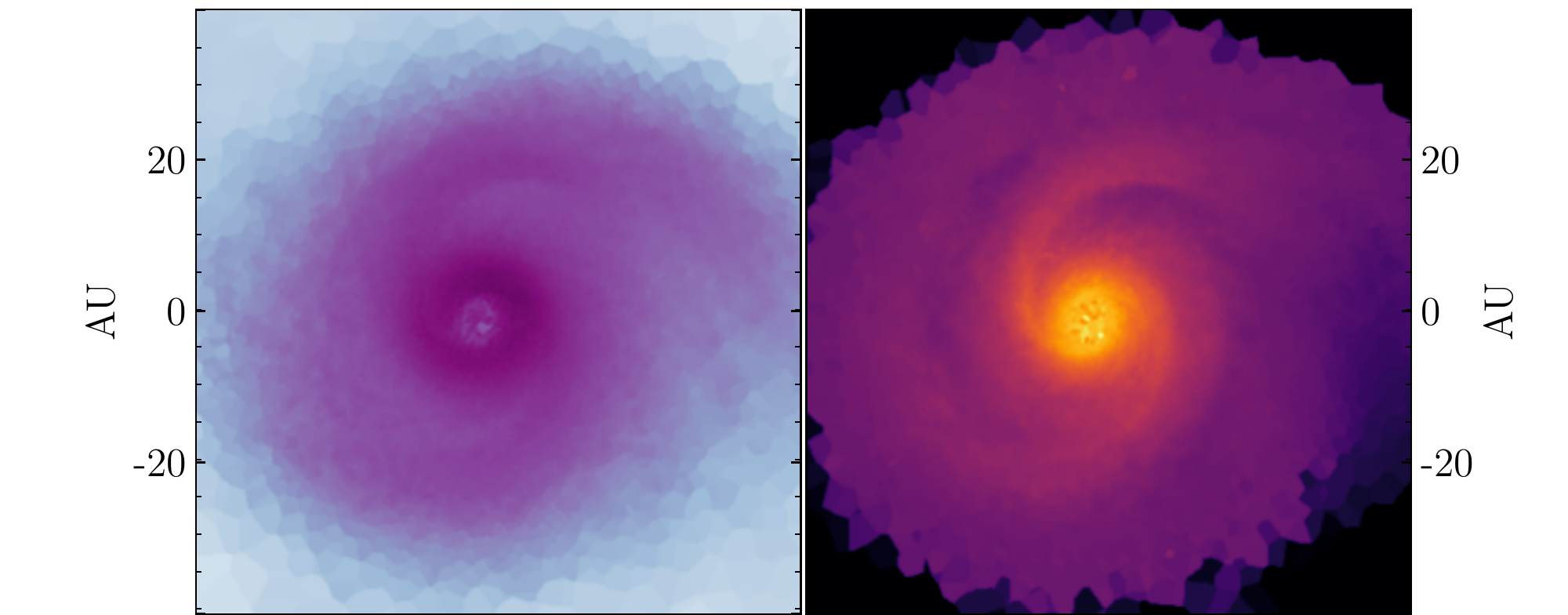}
     \includegraphics[width=16cm, trim=0cm 0.5cm 0cm 12.95cm, clip]{img/disk_model_ilee.pdf}\!\!
     \caption{Face- and edge-on slices of the gas density (left column) and gas temperature (right column) distributions for the RHD the gravitationally unstable model, as described in section \ref{sec:gravitationally_unstable_disk}.}
     \label{fig:ilee_model}
 \end{figure*}

\subsubsection{RHD gravitationally unstable disk model}
\label{sec:gravitationally_unstable_disk}

We also model the formation of a disk in which temperature evolution is followed self-consistently along with the hydrodynamics. For this, we utilise the smoothed particle hydrodynamics (SPH) code \texttt{sphNG} \citep{Bate1995} including the hybrid grey radiation transfer scheme of \citet{Forgan2009} to perform radiation hydrodynamic (RHD) simulations. In this case, material is able to exchange energy via flux-limited diffusion \citep[see e.g.][]{Bodenheimer1990,Whitehouse2004} and the gas is able to cool radiatively according to its local optical depth (estimated from the local gravitational potential, see \citealt{Stamatellos2007}).  We note that this approach does not consider any magnetohydrodynamic processes.

We follow the evolution of a spherical, isothermal cloud of mass 1\,M$_{\odot}$ and radius 2000\,au using $5\times10^{5}$ SPH particles with a background temperature of 5\,K.  After approximately 17 kyr, a rotating disk is formed at the centre of the cloud.  For the rest of this work, we consider the disk physical structure at 18.2 kyr of evolution, approximately 1000\,yr after the disk first forms. At this point, the mass of the disk is 0.3\,M$_{\odot}$ and the sink particle close to the center of the disk has a mass of 0.2\,M$_{\odot}$.  
The disk is gravitationally unstable, with a Toomre $Q$ parameter \citep{Toomre1964} below 1.7 (the critical value for non-axisymmetric perturbations; \citealt{Durisen2007}) at radii larger than 7\,au and a minimum value of $\sim$1.4 at approximately 20\,au (see Fig. \ref{fig:toomre_q}). 
While the disk does not undergo fragmentation during the period of time that we simulate, it does exhibit significant non-axisymmetric structures in the form of two spiral arms within the disk.   
Face-on and edge-on slices of the gas density and gas temperature distributions of the RHD disk model are shown in Fig.~\ref{fig:ilee_model}.



The radial dependence of gas density is similar to that of the MHD model: it reaches a maximum value of $5\times10^{-11}\,$g\,cm$^{-3}$ at around 4~au, then stays roughly constant until 7~au, where it starts to decrease as $r^{-2.6}$ until $\sim$40~au. Between 3 and 13\,au both models have very similar density values and profiles. Beyond this region, the RHD model is denser than the MHD model by a factor of $\sim$2.5.  
Additionally, the density in the RHD disk model is not as smoothly distributed as in the MHD case, even when the two are marginally gravitationally unstable, since the spiral arms are more than twice as dense as the rest of the disk.  

Since the RHD model includes a radiation transfer scheme, the temperature evolution is not only dependant on the hydrodynamics but also on the local radiation field. 
As a result, the coupling between the temperature and density is not as tight as in the MHD model.
The disk temperature is the highest ($T\gtrsim700$\,K) within the inner $\lesssim$5, peaking at the very center with values of around 1000\,K. However, the temperature at the center can be affected by artificial viscosity present in SPH simulations in regions of very high density, as those in the central few au \citep{Bate1995,Forgan2009}. For this reason, when comparing with the observations, we do not focus on the values reached in the very central region which we conservatively consider to be within 10\,au. 
Beyond 10-15 and up to around 30\,au, the overall temperature distribution follows that of the density, with temperature contrasts of 1.5-2 between the spiral arms ($\lesssim$500\,K) and the rest of the disk. 
These outer regions are much hotter ($\sim$200\,K) than the MHD disk model ($\sim$50\,K) by a factor of about 4. The adiabatic equations of state of both models are very similar, but the treatment of the heating and cooling used in this model self-consistently produces higher temperatures at radii of 10-25\,au. This is possibly due to a combination of the high gas densities and kinematics processes that lead to additional compression/shock heating. 

The vertical temperature structure follows a similar gradient to that of the MHD model, increasing from the outer layers to the midplane.

\section{Radiative transfer}
\label{sec:Radiative_transfer}

We have post-processed the disk models previously introduced, by performing radiative transfer (RT) calculations on their density structures, with the publicly available code \texttt{POLARIS}\footnote{\url{http://www1.astrophysik.uni-kiel.de/~polaris}} (v4.06)~\citep{Reissl16}. We use as input the density, velocity and magnetic field distributions at the timesteps of interest. We also consider the gas temperature from the simulations whose effects will be discussed below. 


To facilitate the direct comparison between numerical simulations and real observations, POLARIS supports user-defined input grids, which could be converted from the snapshot of a full 3D-MHD code. In this work we have converted the models from \citet{Zhao2018} (see section~\ref{sec:non-ideal_MHD_model}) in spherical grid structure into the POLARIS spherical grid format and publicly released the code used for the task\footnote{\url{https://github.com/jzamponi/zeus2polaris}}. For the RHD model (see section~\ref{sec:gravitationally_unstable_disk}), the conversion was done from a non grid-based SPH data into a POLARIS Voronoi mesh format, mapping each SPH particle into one cell.

\subsection{Dust grain size distribution and composition}
\label{sec:dust_opacity}

Both the generation of dust temperature fields and the ray-tracing of the dust continuum emission depend strongly on the dust opacity. 
This wavelength-dependent opacity is determined by the composition of the dust mixture and the distribution of grain sizes.
The dust component included in our setup is a mixture of spherical silicates and graphites, meant to resemble the dust composition in the interstellar medium~\citep{MRN} and in protostellar cores~\citep{OssenkopfAndHenning1994}.
The mixture contains a 62.5\% (mass fraction) of silicates and a 37.5\% of graphites.
This mixture has been commonly used in previous work using POLARIS and at similar stages of star-formation~\citep{Reissl16, Valdivia2019, Brunggraeber2019, Kueffmeier2020, Brunngraeber2020}. 
As the temperatures in the disk are mainly above the sublimation temperature of interstellar ices, no ice component has been included in the dust model.  
The graphite component follows the 2/3 - 1/3 relation for cross sections perpendicular and parallel to the incident electric field \citep{DraineAndMalhotra1993}. This means that 37.5\% of the graphite is split into 25\% and 12.5\% of the perpendicular and parallel components, respectively. 
The densities of the silicates and graphites are 3.5\,g\,cm$^{-3}$ and 2.2\,g\,cm$^{-3}$, respectively, leading to a mixture density of 2.896\,g\,cm$^{-3}$. 
For all radiative transfer calculations, we have assumed a gas-to-dust mass ratio of 100. 
 
The dust opacity is calculated from the refractive indices (a.k.a., dielectric constants) using the Wolf \& Voschinnikov approach \citep{WolfAndVoschinnikov2004}, which is in turn an implementation of the commonly used Bohren \& Huffman approach \citep{BohrenAndHuffman} optimized for large size parameters ($x=2\pi a/\lambda$).
We have used the refractive indices for astronomical silicates and graphites from the POLARIS repository, which are based on \citet{DraineAndLee84}, \citet{LaorAndDraine93} and \citet{WeingartnerAndDraine2001}.

POLARIS performs no mixing between the refractive indices of the two dust components (e. g., Bruggeman or Maxwell-Garnet mixing, see \citealt{Ossenkopf1991}), rather it lets the two species coexist in every cell and compute the total opacity as a sum of single opacities weighted by the mass fraction (see e.g.,~\citealt{Das2010}) and removes them from the cell if the dust temperature ever exceeds the sublimation temperature of the material. The sublimation temperatures of silicates and graphites are 1200 and 2100~K, respectively. 

\begin{figure}
    \centering
    \includegraphics[width=\columnwidth, trim=0 1cm 0 0]{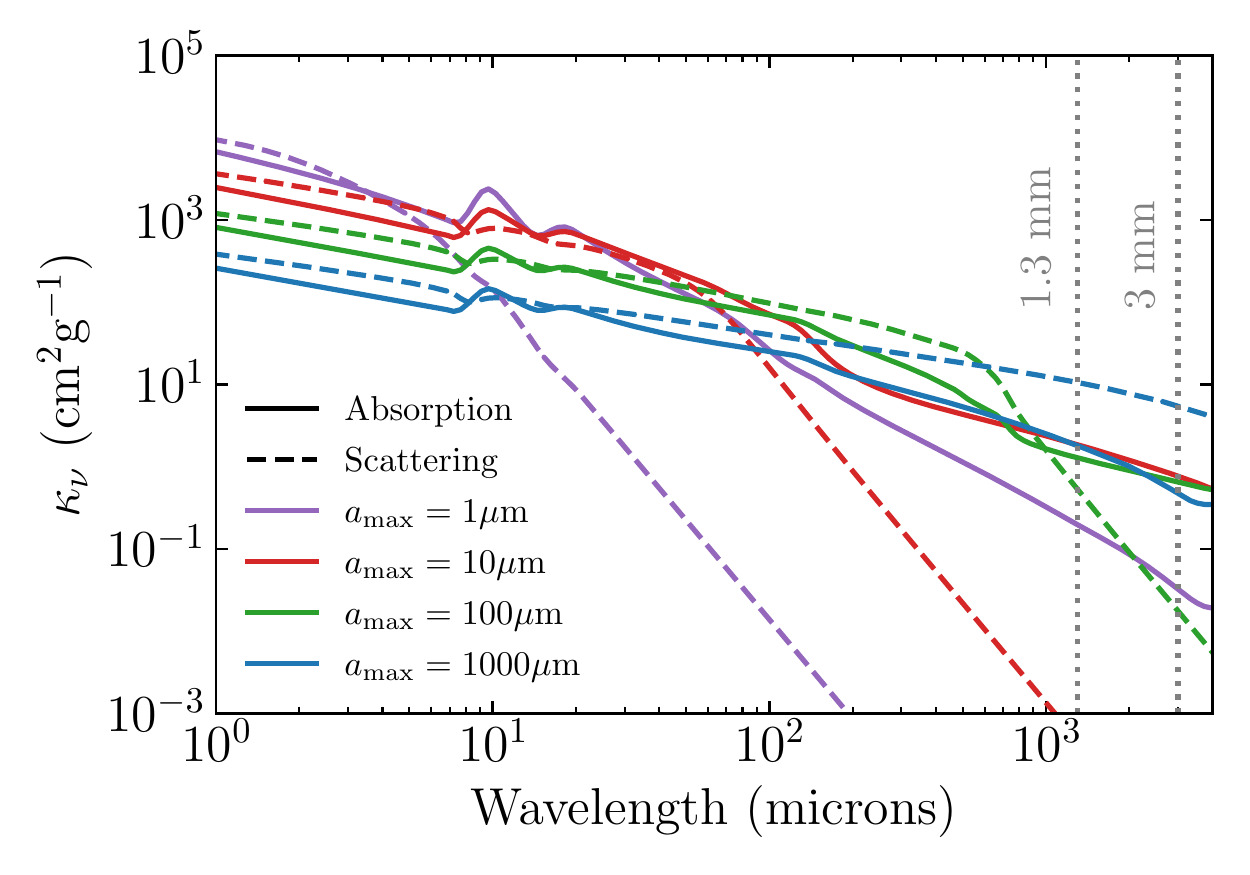}
    \caption{Absorption (solid lines) and scattering (dashed lines) opacities (in units of cm$^2$ per gram of dust) for dust grains of silicate plus graphite composition. All  opacities were obtained assuming a grain size distribution with $q=3.5$ and a minimum grain size of 0.1$\mu$m. Dotted vertical lines at 1.3 and 3 mm indicate the wavelenghts of interest in the present study.}
    \label{fig:dust_opacities}
\end{figure}

The sizes of the dust grains follow a power-law distribution $N(a)\propto a^{-q}$, with $q=3.5$, distributed among 200 logarithmically spaced size bins and 200 logarithmically spaced wavelengths. 
We chose a minimum grain size of $a_{\rm min}=0.1~\mu$m, based on the value used in the disk simulations from \citet{Zhao2018} analyzed here and also consistent with recent simulations of grain growth in dense cores, showing that the population of very small grains is rapidly lost as they are swept out by larger grains \citep{Silsbee2020}. 
This parameter was kept fixed since it has been shown that variations in the lower limit of the grain size distribution do not produce significant effects on the sub-millimeter and millimeter opacities~\citep{Woitke2016}.    
The maximum grain size ($a_{\rm max}$), on the other hand, directly affects the slope of the opacity curve (known as the opacity index $\beta$) at sub-millimeter and millimeter wavelengths. 
The opacity distributions generated by this mixture are shown in Fig.~\ref{fig:dust_opacities} for different $a_{\rm max}$.

Scattering opacities are commonly assumed to be negligible at long wavelengths. This is also consistent with our results for the dust opacity as long as the $a_{\rm max}\leq100~\mu$m. 
In this work we have assumed $a_{\rm max} = 10\,\mu$m. This results in the flux from the disks being mainly produced by thermal emission. 


Since the focus of this study is not to constrain the grain sizes but to test if any of the presented numerical models can be a good fit to the new high-resolution observations of source B, we did not explore all the parameter space for $a_{\rm max}$, which would be analyzed in a future work. 
Considering that there is no unambiguous evidence of large grains at the early Class 0 protostellar stage, as a first step we consider only grains with $a_{\rm max}\leq10\,\mu$m. 
This choice also results in most of the continuum emission coming from thermal dust emission, since the opacity of scattering is very low compared to that of absorption. The dominance of absorption opacities in the millimeter wavelengths is relevant for understanding the flux distributions, however, for the calculation of temperature distributions, the full extinction opacity (scattering plus absorption) at all wavelengths was used.   

Finally, the extinction dust opacities used in this study are $\kappa_{\rm ext}^{\rm 1.3~mm}=1.50$\,cm$^2$\,g$^{-1}$ and $\kappa_{\rm ext}^{\rm 3~mm}=0.58$\,cm$^2$\,g$^{-1}$, as shown in Fig. \ref{fig:dust_opacities}.

\subsection{Dust temperature distributions}
\label{sec:dust_temperature_distributions}

\subsubsection{Dust temperatures from stellar radiation}
\label{sec:dust_heating_from_stellar_radiation}

We have generated 3D distributions of the dust temperature for both disk models, by heating the dust with the radiation from a central star.  
We first consider the dust temperature generated only from stellar radiation, with no inclusion of the gas temperature from the numerical simulations (shown in Fig. \ref{fig:bo_model} and \ref{fig:ilee_model}). 
For this, we placed a point source in the center of the grid, radiating as a black-body (i.e., a protostar) with $R_{\rm star}=3$~R$_{\odot}$ and effective temperature $T_{\rm eff}=4000$~K.
This results in a luminosity of $L_{\rm star}=2.1~{\rm L}_{\odot}$, split among $10^8$ photons ($N_{\rm ph}$).
These stellar parameters are similar to the findings obtained by \citet{Jacobsen2018} using radiative transfer modelling of both protostellar sources (A and B) in IRAS16293-2422. 
They constrained the individual luminosities to $L_{\rm A}=18$\,L$_{\odot}$ and $L_{\rm B}\leq3$\,L$_{\odot}$ for A and B, respectively.

Figure \ref{fig:dust_temperature_profiles} shows the radially averaged midplane dust temperatures for both disk models. The temperature generated by star heating is only relevant within the very central zones in both cases, because the densities in the disks shield the outer regions from stellar radiation. 
In the MHD case, radiative heating overcomes the gas heating only up to 2~au outside the inner boundary and falls below $\sim$60~K at larger radii. 
In the RHD case, it slightly overcomes the gas temperature at sub-au scales and then rapidly falls to a few K at 6~au, showing no contribution to the overall temperature field. 
We have also tested whether these results are dependent or not on the value of the opacities used and found that they hold (i.e., the disk does not heat up significantly beyond a radius of 10 au) even when increasing or decreasing the opacity by an order of magnitude. 
More details on such tests are described in section~\ref{section:opacity_test}.

\subsubsection{Dust temperatures combining stellar radiation and MHD/RHD gas temperatures}
\label{sec:combination_of_dust_temperatures}
In this section, we consider both the gas temperature obtained from the heating mechanisms present in the MHD and RHD simulations $T_{\rm gas}$ (see section \ref{sec:Protostellar_disk_simulations}), and also the dust temperature provided by radiative dust heating $T_{\rm rad}$. 
POLARIS allows to combine these two temperatures in order to account for heating at all scales.
We provide here a brief explanation of the combination process and refer the reader to \citet{Reissl16} for further details.     
In the case of a model with non-zero initial temperature, as for instance, when providing the temperature from the gas, the energy of a cell is considered to be the sum of the energy produced by radiation and that associated to the gas temperature.  
To obtain the energy of the gas temperature, POLARIS compares the new cell energy to the emissivity produced by the gas temperature and solve for the energy offset. 
Once the energy information is updated, the dust temperature is recalculated accordingly.

\begin{figure}
    \centering
    \includegraphics[width=\columnwidth, trim=0.0cm 0.8cm 0.5cm 1.8cm, clip]{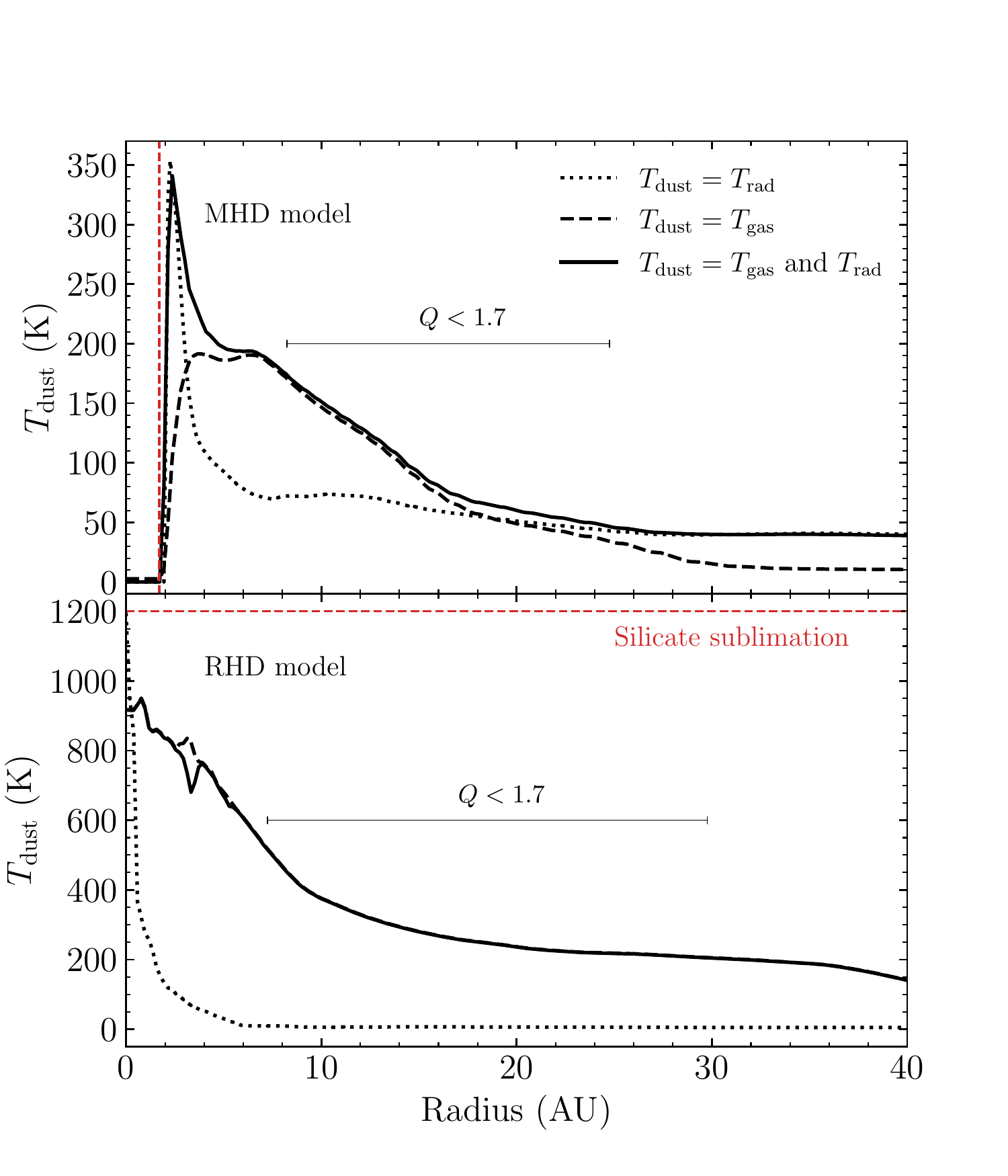}
    \caption{Radially averaged midplane dust temperatures. Temperatures are shown for the MHD disk formation model (top panel) and for the RHD gravitationally unstable disk (bottom panel). In both panels, the dotted line represents the dust temperature produced by radiative heating from the central source only, the dashed line represents the temperature assuming $T_{\rm dust}=T_{\rm gas}$, and the solid line represents a combination of both effects, as described in section~\ref{sec:combination_of_dust_temperatures}. 
    The red dashed line in the upper panel placed at 2\,au, represents the inner boundary in the MHD model, and the red line in the lower panel represents and upper limit for the dust temperature set by the sublimation of silicates. The region where Q<1.7 represents the extensions over which the disk is considered to be gravitationally unstable, as described in Appendix  \ref{appendix:toomre_parameter_for_both_models}.}
    \label{fig:dust_temperature_profiles}
\end{figure}

The gas temperatures in the MHD simulations are given by the EOS while in the RHD simulation the gas is able to cool radiatively according to its local optical depth \citep{Stamatellos2007} and exchange energy via flux-limited diffusion \citep{Bodenheimer1990,Whitehouse2004}. 
The radially averaged profiles for the $T_{\rm dust}= T_{\rm gas}$ and $T_{\rm rad}$, as well as for the $T_{\rm dust}=T_{\rm gas}$ only case for both disk models are shown in Fig.~\ref{fig:dust_temperature_profiles}. 

From this figure, we conclude that for the MHD model, radiation from the protostar dominates within the first 4\,au, but beyond that radius the temperature from the EOS dominates up to around 18\,au.  
At larger radii ($\gtrsim$20\,au) outside disk edge, the protostellar heating dominates again resulting in a fairly constant temperature of 40\,K. 
On the other hand, for the RHD model the protostar radiation is not able to penetrate beyond 1\,au due to the high optical depth.  
Therefore, the disk temperatures at the scales where the spiral arms are located are not dominated by protostellar heating. Instead, the high dust temperature here is a result of the high gas temperature, which is also higher than the one reached by the MHD model. The resultant high dust temperature are due to the efficient thermal coupling at high densities. It is non-trivial to disentangle the sources of extra heating in the RHD model versus the MHD model because they can can come from different sources.  For instance, the models were generated with a different numerical setup, i.e., SPH vs grid-based code, with and without magnetic fields or radiative transfer and slightly different initial conditions. As pointed out by \citet{Whitehouse2006} the thermal evolution of different setups is expected to be different. Moreover, they also discuss that the use of an EOS can underestimate the temperature of the gas surrounding the central highest density. In addition, the spiral arms do add extra compression heating, responsible for the local temperature rise. 

While a comparative modelling study to investigate this is beyond the scope of this work, we can still rule out some possible reasons for the different temperature. For instance, it is not due to the RHD having higher density than the MHD model since the models show different temperatures while covering a similar density range. 
Likewise, the thermal evolution of the central density of the RHD code (shown in Fig. 8 of \citealt{Forgan2009}) and of the MHD code (see Fig. A1 from \citealt{Zhao2018}) is comparable. Also it is not due to the RHD disk being more optically thick, as a setup fully based on a EOS (with no cooling) would represent an optically thick limit.

Based on this analysis, we conclude that an extra source of heating is needed to explain the high brightness temperatures observed and that gravitational instabilities in the disk provide a relatively simple explanation for it.

\subsection{ALMA synthetic observations}
\label{section:ALMA_synthetic_observation}
The ideal intensity distributions produced by POLARIS, in units of Jy~pixel$^{-1}$, needed to be post-processed assuming a certain observing setup and telescope, in order to properly compare them with real observations.
Because our goal is to compare our results to ALMA observations, we performed the synthetic observations using CASA, which provides the right tools to simulate ALMA observations.
We first adjusted the header of our images to match those from the data at 1.3 and 3~mm described in section~\ref{sec:data_calibration}. 
We then used the \texttt{simobserve} task to generate the complex visibilites and finally cleaned the images using the \texttt{tclean} task.
The observing and imaging setups for each frequency are slightly different since the real data was observed in different occasions and with different setups.

Visibilities were first generated at the frequency of 223~GHz (1.3~mm; band 6), with a bandwidth of 0.12~GHz toward the sky coordinates 16~h~32~m~22.63~s ~-24~d~28~m~31.8~s and using the array configuration C43-7 from Cycle 4.
Such a setup achieves a synthesized beam of $0.082"\times0.067"$.
The total observing time was 1.25~hours.
Images were cleaned in multi-frequency synthesis (\texttt{mfs}) mode, with a \texttt{standard} gridder and a \texttt{multiscale} deconvolver. 
The visibilities were cleaned interactively using a \texttt{briggs} weighting scheme with a robust parameter of 0. 
For the observations at 1.3~mm only, we narrowed the range of baselines (\textit{uv}-range) from 120 to 2670 k$\lambda$. 
The 3~mm observations shared the same setup, except that they were centered at the frequency of 99.99~GHz (band 3) with a bandwidth of 2~GHz using the most compact array configuration C43-10 from Cycle 5.
Such a setup achieves a synthesized beam of $0.042"\times0.039"$.
Since the resolutions of the simulated observations are slightly higher than those of the real ALMA observations, all synthetic maps have been smoothed to match the respective resolution of the real observations.

\section{Comparison with observations}
\label{sec:results}

\subsection{Brightness profiles}
\label{sec:comparing_real_and_synthetic_brightness_profiles}
\begin{figure*}
    \centering
    \includegraphics[width=\columnwidth, trim=0cm 1.18cm 0cm 0cm, clip]{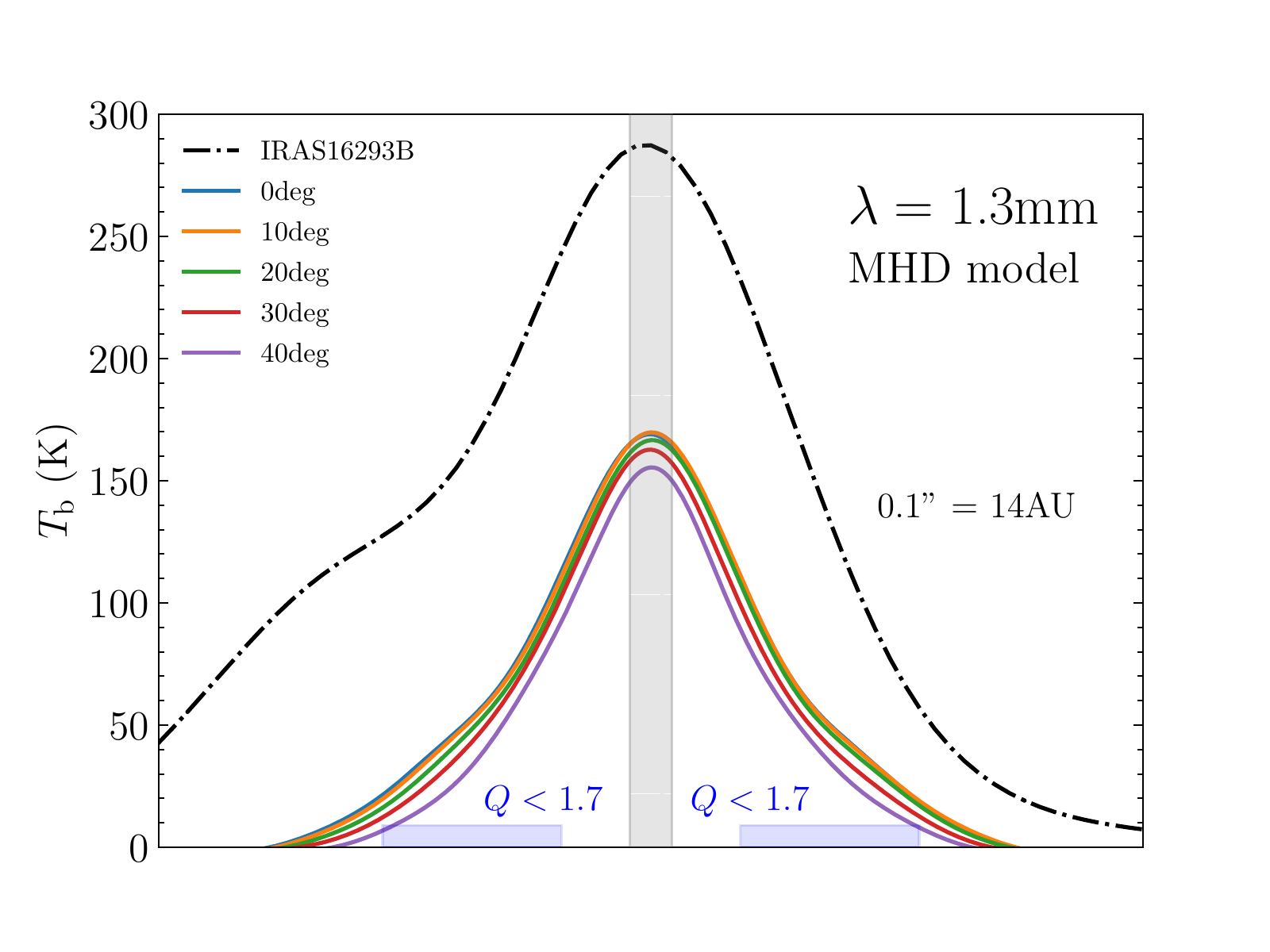}
    \includegraphics[width=\columnwidth, trim=0cm 1.20cm 0cm 0cm, clip]{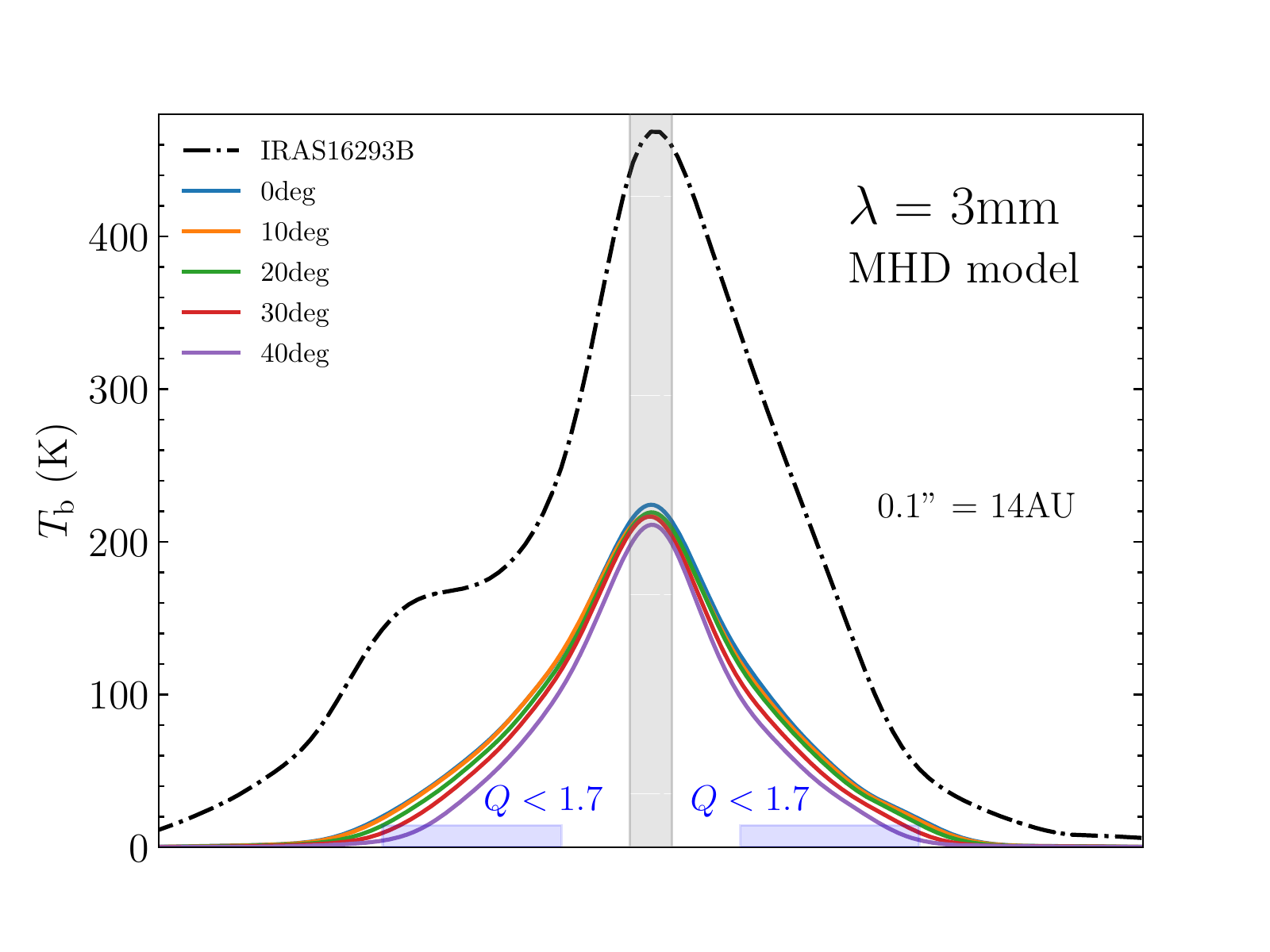} \\
    \includegraphics[width=\columnwidth, trim=0cm 0cm 0cm 1.25cm, clip]{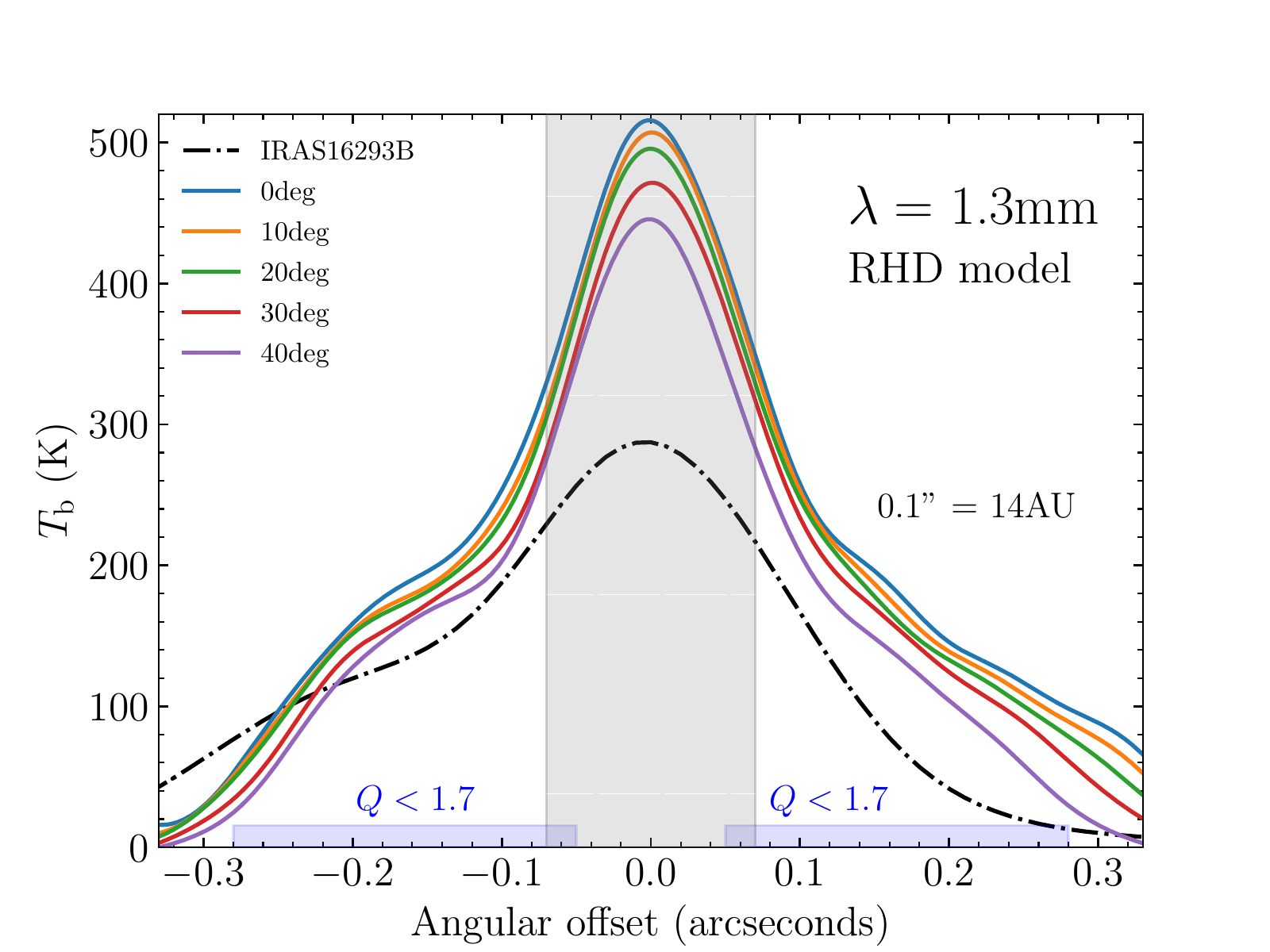}\!
    \includegraphics[width=\columnwidth, trim=0cm 0cm 0cm 1.20cm, clip]{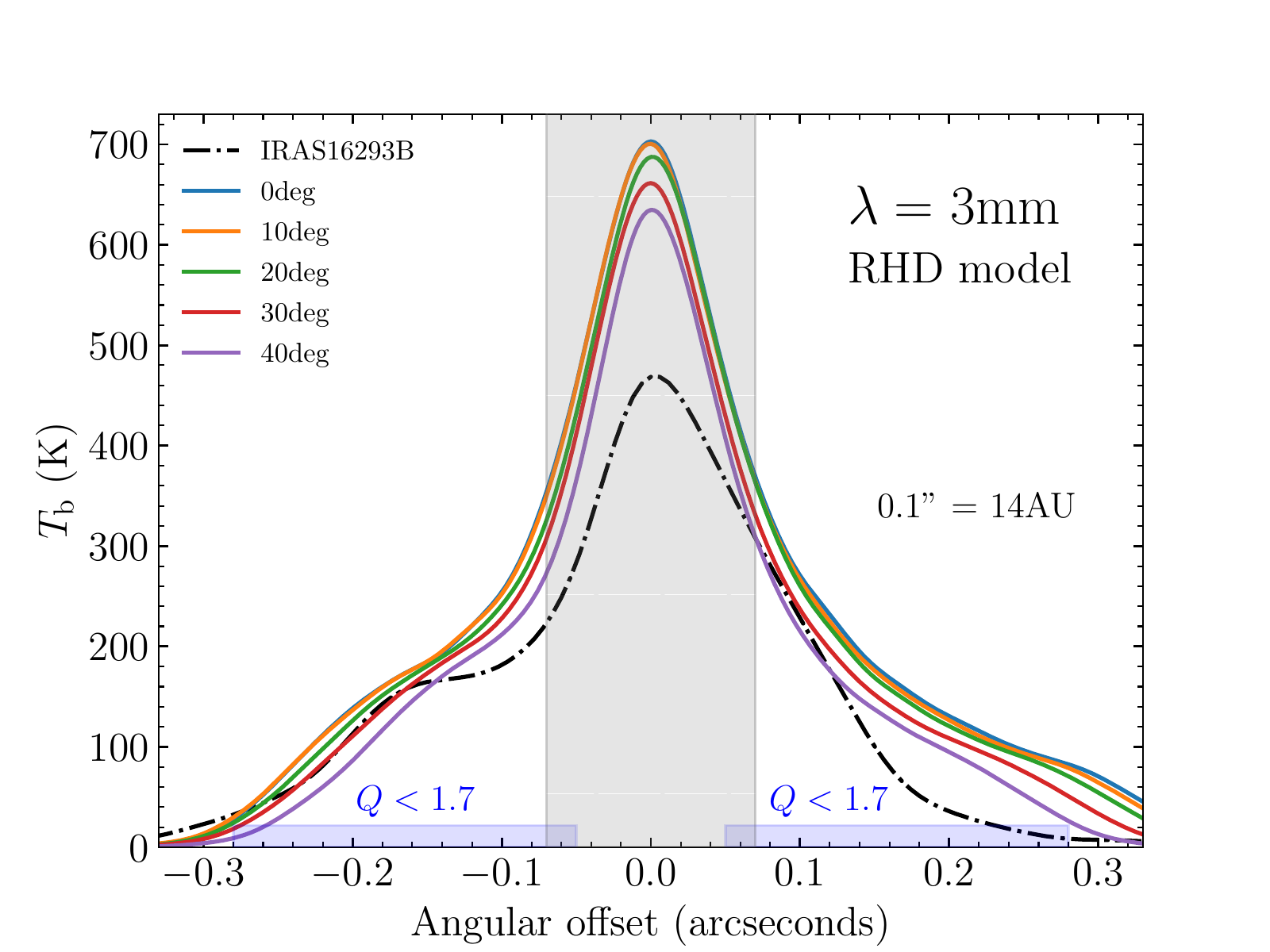}\!
    \caption{Horizontal cuts of brightness temperature distributions along the East-West direction and through the position of the peak flux. Top panels show the results at 1.3 and 3 mm, for the MHD model with $T_{\rm dust} = T_{\rm gas}$ and radiative heating. Lower panels show analogous results for the RHD model, where the disk temperature ($T_{\rm dust}=T_{\rm gas}$) is significantly higher and closer to the observed values. The grey shaded area in the MHD model represents the inner boundary of 2\,au. The grey shaded region in the RHD model represents the inner 10\,au in which the temperatures and therefore the fluxes can be overestimated by numerical high viscosity. The blue shaded region in all panels indicate the extensions over which the disk is considered to be gravitationally unstable, as described in Appendix \ref{appendix:toomre_parameter_for_both_models}. The angular offset is measured with respect to the peak emission in each image.}
    \label{fig:horizontal_cuts}
\end{figure*}

Based on the intensity distributions generated by the radiative transfer calculations, we performed synthetic ALMA observations using the two disk models described in section \ref{sec:Protostellar_disk_simulations} as sources.  
The observations were simulated with CASA (v5.6.2) following the procedure previously described in section~\ref{section:ALMA_synthetic_observation}. 
We analyzed the resulting images by generating cuts of the temperature brightness along the horizontal (east-west) axis and studied their similarities to the brightness profile from the real observations presented in section \ref{sec:observational_results}. 
The resulting profiles are shown in Fig. \ref{fig:horizontal_cuts} at 1.3 (left column) and 3~mm (right column) for the MHD (upper panels) and RHD (lower panels) disk models. 
Profiles are shown for the real observation (semi-dotted black) and for different disk inclinations from 0 (face-on) to 40 degrees (rotated around the north-south direction) and assuming the opacity generated by a size distribution with $a_{\rm max}=10\,\mu$m.
The brightness temperatures are presented as a function of angular offset with respect to the peak emission, i.e., the zero offset corresponds to the peak in each image.  

We have also tested the effects of a dust opacity with $a_{\rm max}=1\mu$m (i.e., $\kappa^{\rm 1.3mm}=0.29$~cm$^2$ g$^{-1}$ and $\kappa^{\rm 3mm}=0.05$~cm$^2$ g$^{-1}$), which is closer to ISM values \citep{MRN}. 
However, the simulated emission with $a_{\max}=1\mu$m fails to reproduce the observed 3\,mm fluxes at scales of $\sim$20\,au from the peak, by a factor of $\sim$80 for the MHD model and $\sim$5 for the RHD model. 
With this smaller $a_{\rm max}$, not only the fluxes are underestimated but also the disk looks much smaller and is incapable of reproducing brightness asymmetries with any of the two models. 
For this reason, the results presented hereafter, are all obtained assuming $a_{\rm max}=10\mu$m. 

Based on the analysis presented in section \ref{sec:dust_temperature_distributions} and shown in Fig.~\ref{fig:dust_temperature_profiles}, for our setup, the heating provided by protostellar radiation is only important when the gas temperature in the disk is relatively low ($T$$\sim$$150$~K), as it is in the case of the MHD model. Therefore we consider both sources of heating (i.e., $T_{\rm dust}=T_{\rm gas} + T_{\rm rad}$) when producing the synthetic emission maps of the MHD model. As the MHD model has a central hole (the inner boundary) with no temperature or density, we filled this hole before the ALMA simulations with a constant flux equal to the average values in the innermost ring (optically thick approximation). We note, however, that the angular extent of this hole is only 0.03 arcseconds and does not contribute significantly to the peak flux.

In the RHD model, the compression/shock heating overcomes radiative heating at almost all radii, and therefore, the results shown for the RHD model considered the dust temperature as provided by the \textbf{gas} temperature only ($T_{\rm dust}=T_{\rm gas}$). Regardless, the inclusion of the protostellar heating in this model would only change the fluxes of the peak, which we are not analyzing since the gas temperatures in this region can be overestimated due to the effect of the artificial viscosity (as explained in section \ref{sec:gravitationally_unstable_disk}). 

Fig.~\ref{fig:horizontal_cuts} shows that the peak brightness generated by the MHD model underestimates the values observed with ALMA by factors of 1.7 and 2, at 1.3 and 3~mm, respectively, when the disk is observed face-on. 

We have also tested the effect of increasing the luminosity of the central protostar to 5 and 20~L$_{\odot}$, but this is still underestimating the fluxes even in the most luminous case, in agreement with the above-mentioned finding of rapid shielding from stellar flux at smaller radii. From all this parameter study, we conclude that the temperatures in the MHD disk are not capable of reproducing fluxes as high as those observed towards source B in these high-resolution observations, even when combining the temperature from both heating sources.


When considering both models beyond 10 au up to 40 au, the fluxes of the RHD model are much closer to the observed values than the case for the MHD disk. This is indeed the spatial scale at which the Toomre parameter is below 1.7 and the spiral arms are found in the RHD model. The fraction of underestimation from the MHD model at these scales (measured at a left-hand-side of the peak) is around 3 and 3.5 at 1.3 and 3\,mm, respectively, while for the RHD model, the corresponding overestimation fractions are only 1.4 and 1.1 (measured at 0.14 arcseconds or $\sim$20 au). As it can be seen, the match between simulated and real fluxes is better at the higher resolution observations, namely at 3\,mm.

Our results also indicate the presence of asymmetric horizontal brightness profiles in the RHD disk model, similar to what is observed in Fig.~\ref{fig:IRAS16293B_observations}, and follow a similar behavior to the left and right of the peak, i.e., a rapid decrease in intensity until 0.1" offset (14~au; $T_{\rm b}\sim180$~K) followed by a constant drop on the right and a two-phase drop (a.k.a, wing) on the left. 
Both features are best matched at the highest resolution observations taken at 3\,mm. 
The asymmetric feature of the observations lies within offsets $\sim$-0.07" and -0.3".  
These are exactly the spatial scales at which the spiral arms are found in the RHD disk model ($\sim$10-40\,au; see Fig.~\ref{fig:ilee_model}), which supports that this type of asymmetry could be produced by the presence of spiral arms. 
However, this is only one possibility for the origin of the asymmetry and we acknowledge that other processes such as assymetric accretion from the envelope onto the disk, based on the dectection of infalling material \citep{Pineda2012}, may also be taking place.  
Further parameter space explorations are needed to rule-out any of the possible scenarios. 

We also find that the more inclined disks produce lower fluxes than the face-on case. This can be explained by the higher optical depth obtained when the Line Of Sight (LOS) goes through more disk material, as it is the case for an inclined disk.   

Since the RHD model, over scales of 10 to 40\,au, appears to reproduce well the observed fluxes, we conclude that the high disk temperatures reached by the RHD gravitationally unstable model represent a good match to the ALMA observations of IRAS16293-2422 B. 
This suggests that the gas dynamics within the disk likely play a significant role in heating the disk. 

\subsection{Synthetic maps and spectral index of the RHD gravitationally unstable model}
\label{sec:simulated_observations_grav-unstable_model}
In this section, we present the synthetic observations at 1.3 and 3\,mm along with its spectral index, for the best match to our observations, i.e., the RHD model. 
These results are shown in Fig.~\ref{fig:simulated_observations}. 
We also provide the results for the MHD disk model in Appendix~\ref{appendix:spectral_index_bo_model}.  
Similar to the real observations, the 3\,mm map has been smoothed with the beam of the 1.3~mm observation before computing the spectral index. 
We present the results obtained by a dust grain size distribution with $a_{\rm max}=10\,\mu$m and for the face-on projection.  
At both wavelengths, the peak is approximately centered, although the extended structure is not circular or entirely symmetric.  
The substructures present in the disk, such as spiral arms (see Fig.\ref{fig:ilee_model}) are not clearly seen at 1.3\,mm ($0.11"\times0.07"$ resolution), similar to the case with the real ALMA observations. 
On the other hand, the 3\,mm observations, with their higher angular resolution and lower optical depth, display better the presence of spiral arms. 

In analogy to Fig.~\ref{fig:IRAS16293B_observations}, we present the derived spectral index between 1.3 and 3~mm (223 and 100~GHz) for the same disk model, with contours ranging from 3, 2 to 1.7.  
Our results indicate a drop to around $\alpha\lesssim1.8$ at the very center of the source, similar to what is observed in the ALMA observations (see Fig. \ref{fig:IRAS16293B_observations}). 
This result is not only found in the case of the RHD disk model. 
In Fig.~\ref{appendix:spectral_index_bo_model} we show the results for the MHD disk model in a similar layout as Figs.\ref{fig:IRAS16293B_observations} and \ref{fig:simulated_observations}. 
The spectral index found in the MHD case is consistent with both the ALMA observations and the RHD model, with a tendency to decrease towards the center, reaching values as low as 1.7. 
In section \ref{sec:are_early_disks_hotter_in_the_midplane} we discuss the origin of such a low $\alpha$ values in the observations and both models.

\begin{figure*}
    \centering
    \includegraphics[width=18cm, trim=1cm 0 1cm 0cm, clip]{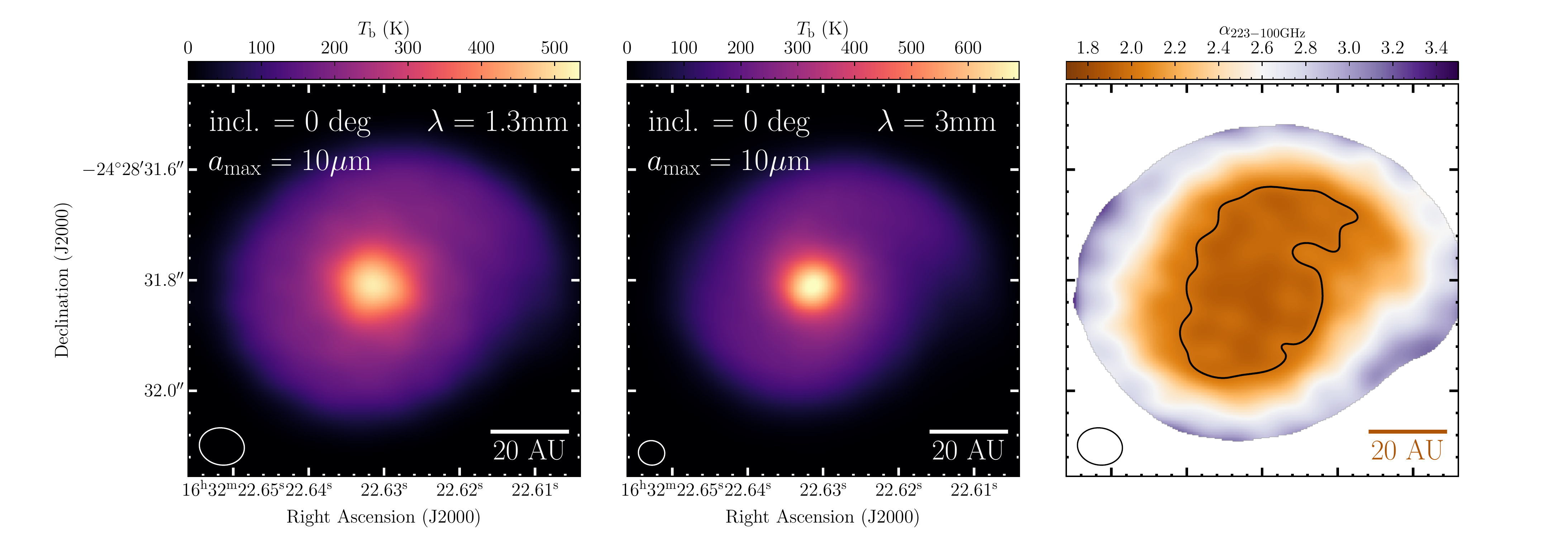}
    \caption{Synthetic brightness temperature maps at 1.3 and 3~mm along with the spectral index for the gravitationally unstable RHD model. The spectral index map reproduces well the trends seen in the observed map shown in Fig. \ref{fig:IRAS16293B_observations}. The contour for the spectral index indicates the $\alpha=2$ region.}
    \label{fig:simulated_observations}
\end{figure*}

\section{Discussion}
\label{sec:discussion}

\subsection{Are early disks hotter towards the midplane?}
\label{sec:are_early_disks_hotter_in_the_midplane}

The tendency for the spectral index $\alpha$ to decrease at the center of the disks found in the ALMA real and synthetic observations is an interesting observational feature that can provide us with information about the internal structure of early disks. 
The spectral index not only provides information on the slope of the SED but also indicates the presence of temperature gradients along the LOS.  
The dust opacity for a given grain size distribution follows a power-law function of wavelength and is therefore higher at 1.3 than at 3~mm. 
If the disk becomes optically thick along the LOS at both wavelengths, each wavelength will trace a different temperature layer within the disk, with the longer wavelength tracing the more deeply embedded layer. 
In the case of a positive temperature gradient towards the denser inner regions, which is the opposite of what's expected for more evolved passive disks \citep[see e.g., ][]{KenyonAndHartmann1987,Chiang&Goldreich1997,ArmitageBook2009,ArmitageReview2011,Dartois2003,Dullemond2007, Kama2009,Lizano2016,Tapia&Lizano2017,Paneque2021} where the dust temperature is mainly determined by radiative heating, lower frequency observations will trace hotter layers than higher frequency observations, leading to $\alpha\lesssim2$ values as described in Appendix~\ref{appendix:spectral_index}.  

\begin{figure}
    \centering
    \includegraphics[width=0.8\columnwidth, trim=3cm 1cm 3cm 0, clip]{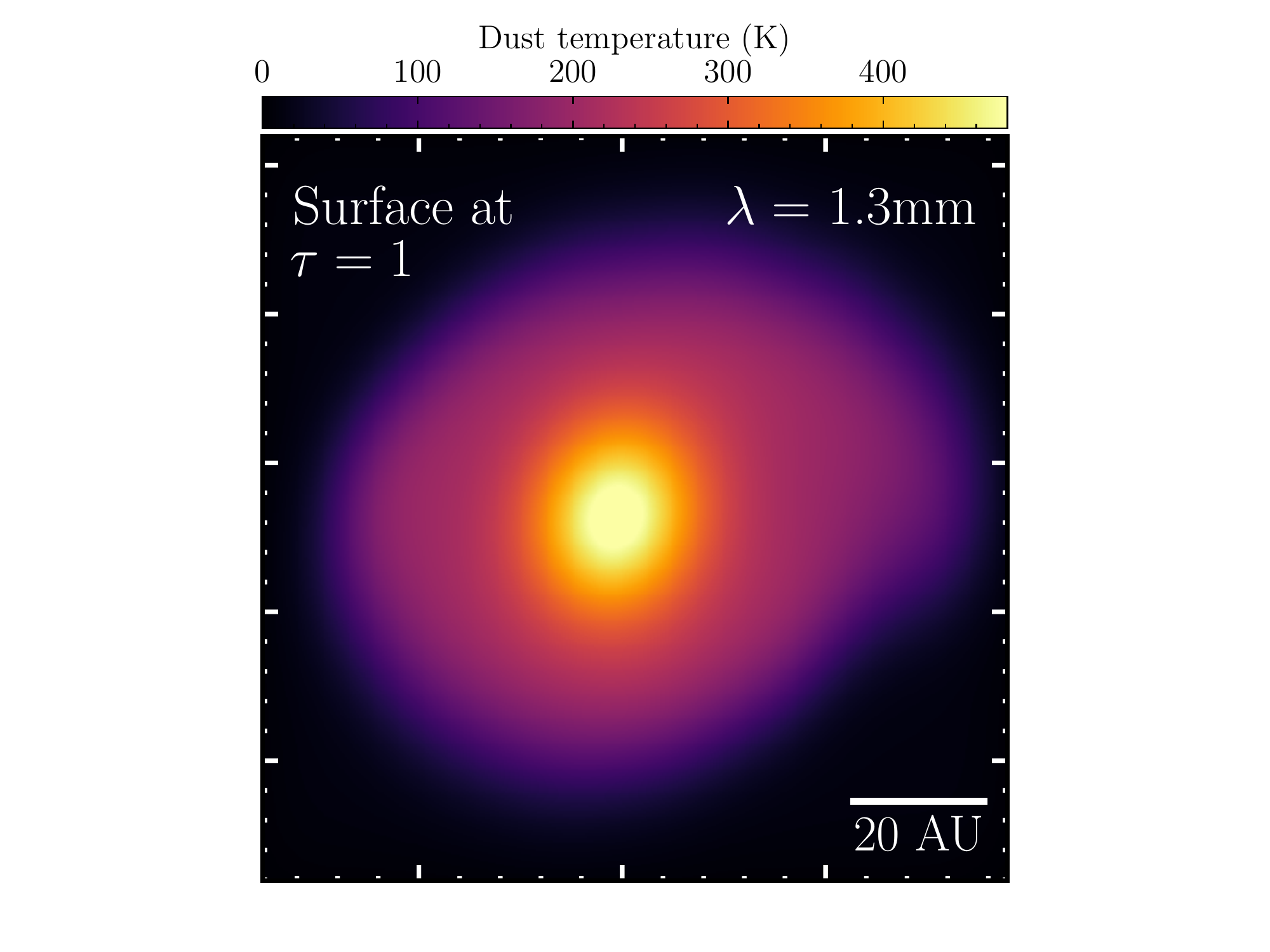}
    \includegraphics[width=0.8\columnwidth, trim=3cm 1cm 3cm 0, clip]{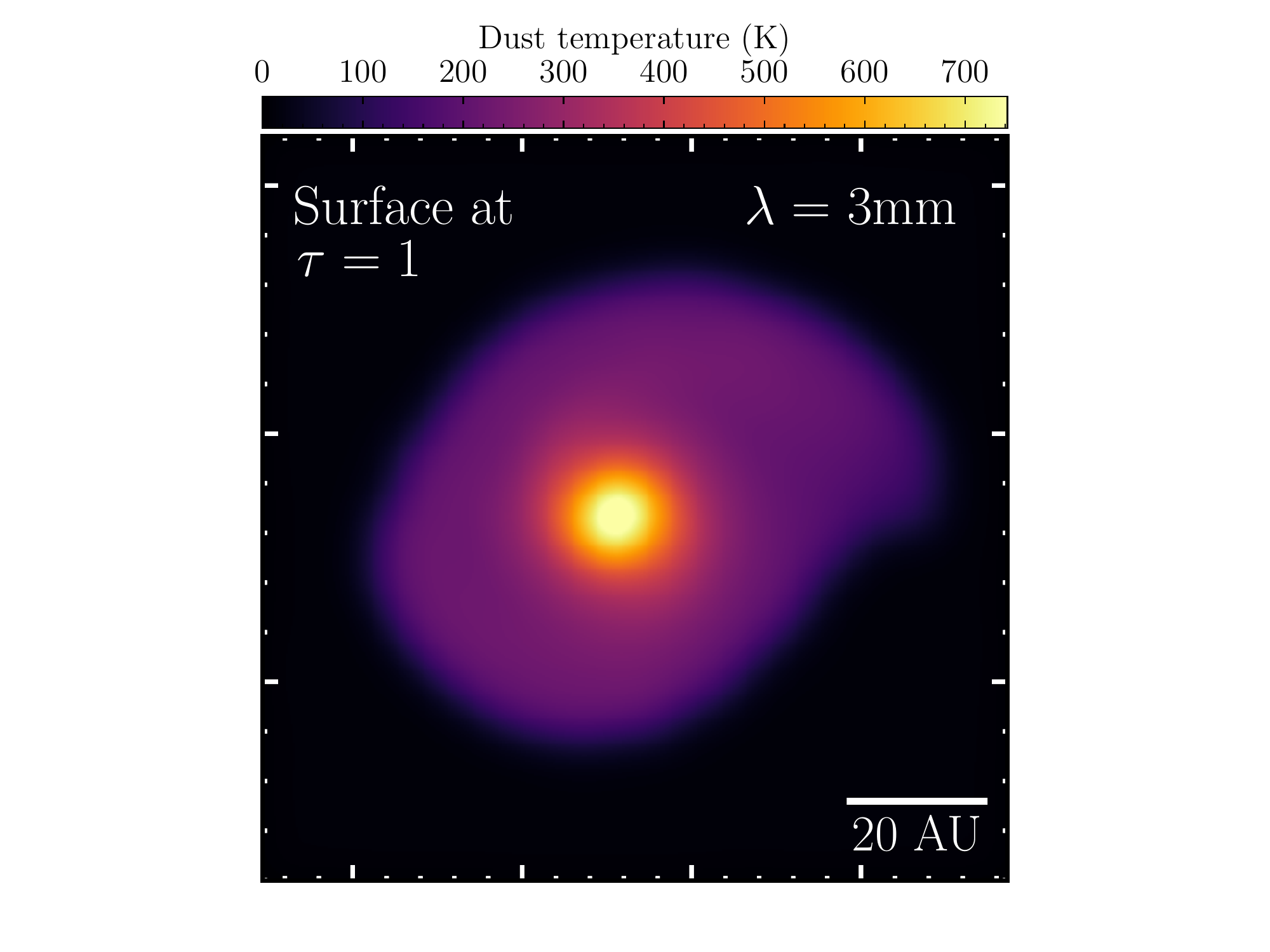}
    \caption{2D face-on distribution of the dust temperature ($T_{\rm dust}=T_{\rm gas}$) from the RHD gravitationally unstable model, at the surface where the optical depth becomes unity. The temperature matches well the brightness temperature distributions shown in Fig. \ref{fig:simulated_observations}, demonstrating that the two wavelenghts are indeed tracing different layers. A 3D view of such surfaces is shown in Fig.~\ref{fig:temp_at_tau1_3D}.}
    \label{fig:temp_at_tau1_2D}
\end{figure}

In principle, the brightness temperature resulting from our synthetic observations should trace the dust temperature of the layer within the disk at which the optical depth $\tau$ becomes unity \citep[see also][]{Evans2017}. 
To prove our point, we show in Fig.~\ref{fig:temp_at_tau1_2D} the 2D dust temperature distribution of the RHD disk at the $\tau=1$ surface for both wavelengths, convolved with the corresponding beams of the ALMA observations. 
Such temperature structures match well those obtained from the radiative transfer calculations indicating that indeed both wavelengths trace two different temperature distributions or layers from different depths within the disk. 

\begin{figure*}
    \centering
    \includegraphics[width=\textwidth, trim=0 0cm 0cm 0cm, clip]{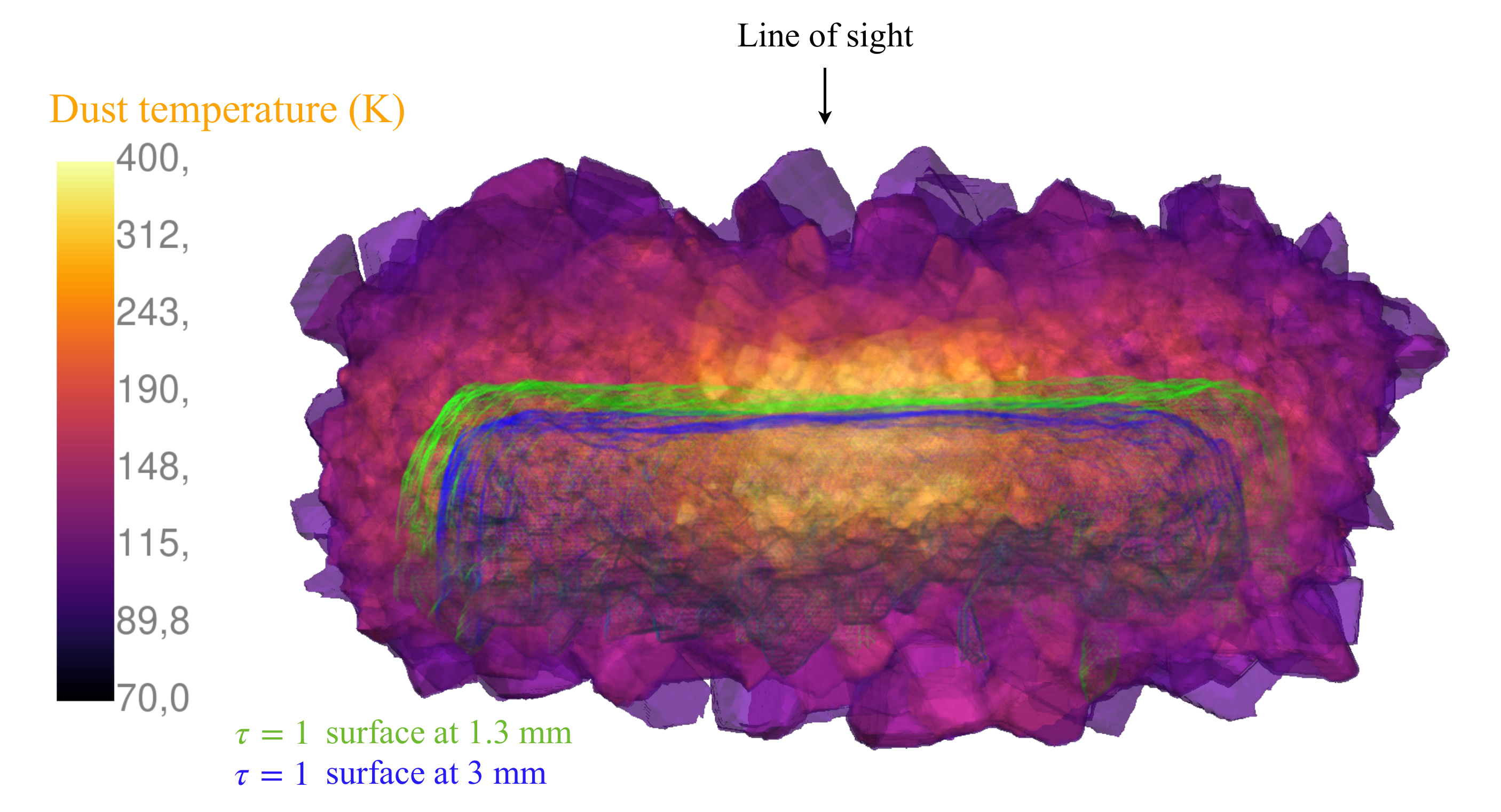}
    \caption{3D rendering of the dust temperature ($T_{\rm dust}=T_{\rm gas}$) for the RHD disk model in edge-on projection. The green and blue surfaces represent the $\tau=1$ layers at 1.3 and 3~mm, respectively. From the observer's point of view, the two optically thick surfaces lie at different depths within the disk and therefore trace dust layers at two distinct temperatures.}
    \label{fig:temp_at_tau1_3D}
\end{figure*}

In Fig.~\ref{fig:temp_at_tau1_3D} we present a 3D rendering of the dust temperature field of the RHD model, represented by the black-to-yellow colored isocontours and we include the $\tau=1$ surfaces at 1.3 and 3~mm in green and blue, respectively. 
Assuming the observer is placed far above the top surface of the disk, this illustration shows the regions of the disk that the observations at 1.3 and 3\,mm have access to.  
We obtain a similar result for the MHD model, which fluxes although too low for reproducing source B observations, can match observations towards other less luminous Class 0 disks. 
Hence, our results from these two models indicate that positive temperature gradients towards the inner denser regions can naturally explain the observations of low spectral index towards young disk sources, especially those with values below 2. 
These low spectral indices are usually interpreted as a sign of early grain growth, but early disks being warmer in their interior as compared with the later Class II disks is an alternative scenario, one that better matches with the theoretical predictions showed in this work. 
This scenario has also been suggested in previous works in which analytical temperature and density profiles were assumed \citep{Galvan-Madrid2018,Lin2021}. 

Recent constraints from molecular line observations have also shown that embedded disks are warmer than more evolved protoplanetary disks.   
For instance, the $^{13}$CO and C$^{18}$O observations towards the nearly edge-on disk (85$^{\circ}$; \citealt{Ohashi1997,Tobin2008,Oya2015}) around the Class 0 protostar L1527, suggests that the midplane temperature is at least $\gtrsim$25\,K \citep{vantHoff2018}, which is contrary to the case of CO depletion in the midplane of Class II disks \citep{Guilloteau2016,Dutrey2017}. 
Warm disks have also been observed in several other disks in Taurus \citep{vantHoff2020c}. 

The presence of a positive temperature gradient at small radii seems to be also the case in the evolved and gravitationally unstable \citep{Dong2016} FU Ori disk \citep{Lizano2016,Liu2019b,Labdon2021}. 
The high-resolution (0.1-10~au) observations reported by \citet{Liu2019b} show the inner parts of the disk in FU Ori with temperatures above 700~K at radii $\lesssim3$~au. 
Their interpretation of the results suggests that the temperature in the inner 10~au is determined by the heating from gas kinematics and that radiative heating becomes important at larger radii, however, they acknowledge not having considered extra sources of heating (e.g., shocks and adiabatic compression) which could also dominate the viscosity as the main heating mechanism.

\subsection{Are self-gravitating disks at the origin of some hot corinos?}
\label{section:are_self-gravitating_disks_at_the_origin_of_hot_corinos}
The high temperatures ($>$100\,K) of the IRAS 16293-2422 B disk implies that most of the volatile material in the dust icy mantles is thermally evaporated, including water and complex organic molecules accumulated during the prestellar phase. 
This process is likely at the origin of the rich chemistry observed by recent spectral line surveys \citep[e.g.][]{Jorgensen2018}. 
It is then tempting to suggest that self-gravitating disks provide the conditions for the detections of the so-called `hot corinos', hot regions nearby Class 0 sources, rich in complex organic molecules \citep[e.g.][]{Cazaux2003}, since it has been shown that gravitational instabilities in a disk directly affects their chemical evolution \citep{Ilee2011, Evans2015, Quenard2018}.  
It is interesting to note that a self-gravitating disk with spiral arms has also been suggested to be present in the Class 0 hot corino HH 212 \citep{Tobin2020a, Lin2021, Lee2021}. 
Additionally, another potentially gravitationally unstable disk in Orion A, HOPS-87 \citep{Tobin2020a} has been recently detected in methanol emission \citep{Hsu2020}, providing additional evidence for the presence of self-gravitating disks in hot corino sources. 
Studies similar to those presented here should be carried out in the currently known hot corinos to test our suggestion.

We finally note that within the central 20 au, in particular along the spiral arms of the self-gravitating disk where shocks are present (see Fig. \ref{fig:dust_temperature_profiles}), the temperature can exceed 300 K (as also found by \citealt{BoleyAndDurisen2008}), allowing efficient sublimation of carbonaceous grains \citep[e.g.][]{vantHoff2020b}. 
This is expected to affect the dust properties (as the important carbonaceous component will be locally depleted) as well as the gas phase chemical composition, due to the local increase of carbon atoms and hydrocarbons. 
This prediction could be tested with high angular resolution multiwavelength observations of the dust continuum emission (to measure possible opacity changes) as well as of C-bearing species (to measure possible increase of the C/O ratio at these locations) at low frequencies (to avoid dust opacity problems; e.g. \citealt{DeSimone2020a}).

\subsection{Radiative heating under different assumptions}
\label{section:opacity_test}
The complexity of a full 3D HD simulation requires that the inclusion of radiation transport should be somehow simplified as compared to a full 3D radiative transfer calculation like those performed with POLARIS. 
In particular, these two radiative transfer schemes differ in the frequency dependency of the opacities \citep{Boley2006}. 
The post-processing done in this paper is performed considering wavelength-dependent stellar fluxes and dust opacities.  
On the other hand, the RHD simulation makes use of a frequency-averaged opacity, the so-called Rosseland mean opacity, which changes to account for different temperature and density conditions (see Fig. 3 from \citealt{Forgan2009}). 

In section~\ref{sec:dust_temperature_distributions} we discuss about the main heating mechanism determining the overall disk temperature in our MHD and RHD models. 
The results from the radiative equilibrium calculations suggest that protostellar radiation is not able to raise the disk temperature significantly beyond $\sim$1~au and therefore the disk temperature is dominated by dynamical processes. 
However, here we want to check if these results also hold when assuming different opacities. In particular, constant opacities as those used in the RHD model. In order to test this effect, we ran radiative equilibrium calculations using roughly constant opacities between 0.5\,cm$^2$\,g$^{-1}$ to 800\,cm$^2$\,g$^{-1}$, which covers the opacities range used in the RHD model \citep{Forgan2009}, for temperatures less than 10$^3$ K. From these tests, we concluded that still protostellar heating alone is not sufficient to heat the disk to temperatures that can explain the observations. In addition, we performed these tests together with increasing the luminosity of the central source up to 20 L$_{\odot}$ for the MHD disk, which is the less dense within the inner 5 au among the two models. 
We found that the temperature obtained from radiative heating alone, peaks at $\sim$400\,K within 1\,au and falls rapidly to 150\,K at 5\,au, beyond which the profile closely follows that of the gas temperature (dashed curve from the upper panel in Fig.~\ref{fig:dust_temperature_profiles}).  
This is obtained when using 20 L$_{\odot}$ and 80 \,cm$^2$\,g$^{-1}$. This luminosity is close to the luminosity values inferred for the entire triple system IRAS 16293, and thus, represents an upper limit. As discussed in section~\ref{sec:comparing_real_and_synthetic_brightness_profiles}, this temperature profile cannot explain the observations. Therefore, heating mechanisms due to dynamical processes are needed to explain the observations.

\subsection{Comparison to spectral index in more evolved gravitationally unstable disks}
\label{section:comparison_to_elias2-27}

\begin{figure}
     \centering
     \includegraphics[width=1.0\columnwidth, trim=1.8cm 0 1.8cm 0, clip]{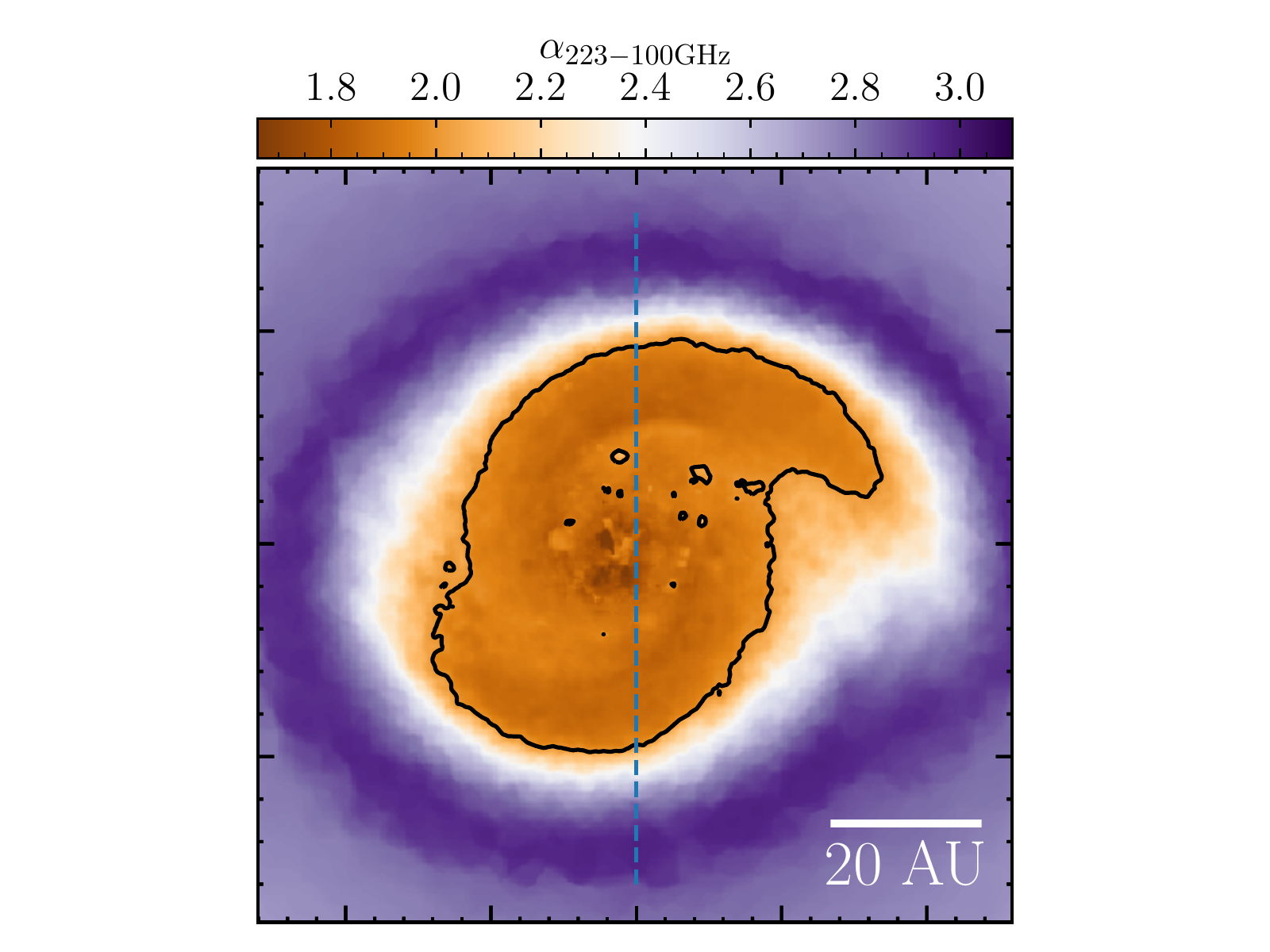} \\
     \includegraphics[width=1.0\columnwidth, trim=0 0 0 0, clip]{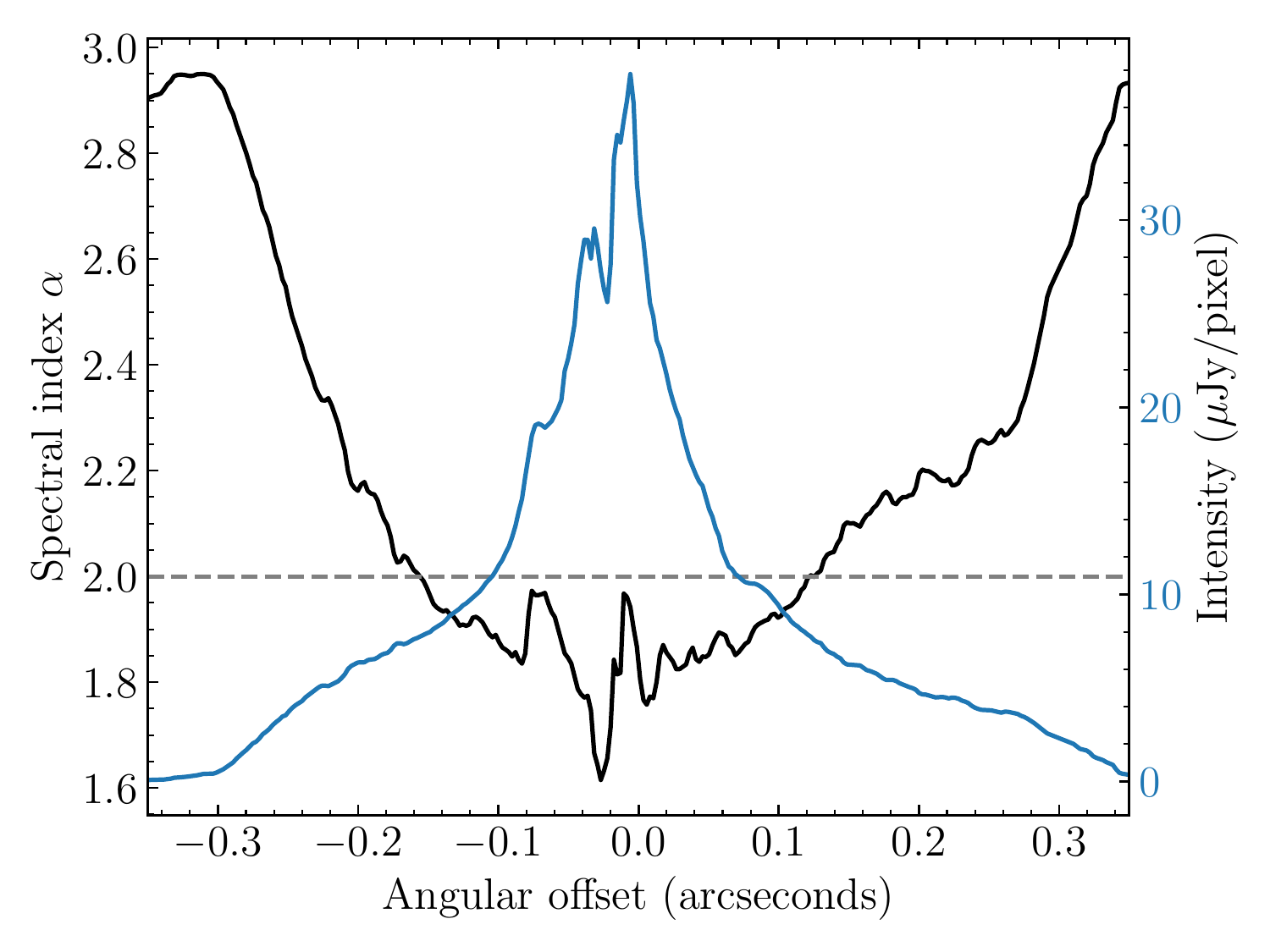}
     
     \caption{(top panel) Spectral index map for the RHD model disk shown in Fig.~\ref{fig:simulated_observations} before performing the ALMA simulation, i.e., at full spatial resolution. The black contour indicates the region with $\alpha=2$ and the vertical dashed line indicates the cut region for the bottom panel. (bottom panel) The black line represents a vertical cut of the spectral index through the center of the image, for the face-on RHD model shown above, and the blue line represents the brightness distribution along the same cut for the 3\,mm simulated observation, without convolution with the ALMA beam. The angular offset is measured with respect to the center of the map.}
     \label{fig:spectral_index_no_convolution_HD_model}
\end{figure}

Recently, the disk mass and milimetric spectral index for the Class II gravitationally unstable disk Elias 2-27 have been reported by 
\citet{Paneque2021} and \citet{Veronesi2021}. This protoplanetary disk shows spiral arms on $\gtrsim$200 au scales, at which location the brightness temperature at 0.89, 1.3 and 3 mm is below 10 K, one order of magnitude lower than the observations of the Class 0 disk in this work. The derived disk mass is about 0.1 M$_{\odot}$, this is factor of 3 lower than the RHD disk model presented in this work. The spectral index calculated using ALMA observations for Elias 2-27 at 0.89, 1.3 and 3 mm falls below 2.0 within $\lesssim$50 au radius. As they also pointed out by the authors, such a low spectral index could be due to dust scattering effects \citep{Liu2019a}, but also due to relatively low dust temperature \citep{Sierra2020}. Alternatively, the heating mechanisms and the radiative transfer effects introduced in our present work may also partly contribute to that result as the inner region is expected to have higher densities. Given that Elias 2-27 is consistent with being gravitationally unstable, the gas kinematics may also play a non-negligible (e.g., as compared with protostellar irradiation) role heating the inner $\sim$50 au of the disk, in particular, in the presence of spiral arms (c.f., \citealt{Dong2016}). As a demonstration, Fig. \ref{fig:spectral_index_no_convolution_HD_model} shows the synthesized spectral index maps of our RHD model at the ordinary spatial resolution.
It appears that the spectral indices can be low over a relatively spatially extended area, and are particularly low at the locations of the spiral arms. 
How important is this effect in Elias 2-27 can be discerned by constraining the dust brightness temperature and dust temperature with future high angular resolution and multi-frequency observations (e.g., including measurements at $\sim$500 and $\sim$700 GHz).
The mechanisms we mentioned are not mutually exclusive.
However, in the case that the low spectral indices are mainly due to low dust temperatures, the 0.89-1.3 mm spectral indices should be positively correlated with dust temperature and may not have a specific relation with the spiral structures.
In the case that it is mainly due to non-radiative heating (e.g., compression or shocks), the 0.89-1.3 mm spectral index is expected to be anti-correlated with the dust temperature, and we should expect the spiral arms to be the local minimum of the 0.89-1.3 mm spectral indices.

\subsection{The origin of the observed asymmetry}

As discussed in Section~\ref{sec:continuum_observations}, there is evidence also from continuum observations at $\sim33-43$ GHz, molecular line distribution and kinematics showing that the asymmetry might be real and not a product of optical depth in a highly inclined disk \citep{Rodriguez2005,Pineda2012,Zapata2013,Oya2018,Hernandez-Gomez2019a}.

Spiral arms are a natural feature in gravitationally unstable disks \citep{KratterAndLodato2016}. 
Depending on the physical conditions, they can be either symmetric or highly asymmetric structures. 
Regardless of their shape, a disk containing spiral arms will show brightness profiles that depart from a symmetric Gaussian profile, as the ones observed in Fig. \ref{fig:horizontal_cuts}. 
The exact shape of such a brightness profile depends strongly on the shape and position of the arms. Even if the arms are symmetric, they will not necessarily produce a flux pattern with symmetric wings around the peak, as we showed in the case of the RHD model in Fig.~\ref{fig:horizontal_cuts}. 
Moreover, the position of the peak in a gravitationally unstable disk must not necessarily be centered within the disk. 
This shift of the peak from the center can arise by considering highly asymmetric spiral arms, unlike those shown by our RHD model. For instance, recent non-ideal MHD simulations presented by \citet{Coutens2020} showed a spiral arm pattern, produced by considering the collapse of a non-rotating 1\,M$_{\odot}$ core initialized with turbulence. 
Such an asymmetric structure, likely produced by the effect of the initial turbulent velocity field and magnetic field could explain the shift of the peak observed toward source B. 
This would also suggest that source B might have formed out of a core with initial turbulence or perturbation instead of solid-body rotation, which would be in agreement with the different orientations of the structures and rotation axes for sources A and B \citep{Pineda2012, Maureira2020}, similar to what has been suggested for other close Class 0 multiple systems \citep{Hara2021}. 
 
Another possibility not involving spiral arms would be that the position of the continuum is not tracing the position of the protostar.  For instance, in the distribution of the spectral index shown in Fig.~\ref{fig:IRAS16293B_observations}, the minimum value is fairly close to the center of the overall structure, while the peak of the dust continuum emission is not. This difference may be due to the continuum peak tracing instead the position of a hot structure close to the protostar and not the protostar itself, which would also be supported by the assymetric distribution of the complex organic molecule emission \citep{Calcutt2018a,Calcutt2018b,Manigand2020,Manigand2021}. This assymetry could also be associated with asymmetric accretion from the envelope onto the disk, based on the previous detection of infalling material in source B \citep{Pineda2012,Zapata2013}. Follow-up observations with similarly high angular resolutions at longer wavelengths (e.g., VLA observations) would be helpful to further assess the origin of the asymmetry.

\subsection{Mass estimates under the optically thin approximation}
\label{sec:disk_mass}

The masses of observed disks are generally calculated assuming that the dust emission is optically thin.
However, as previously discussed, the model that best reproduces our observations is highly optically thick. 
Therefore, similar to the analysis done by \citet{Evans2017}, we investigate here how much are the masses under- or overestimated in the case of embedded disks, when using this common approach. 

\begin{figure}
 \centering
 \includegraphics[width=1.0\columnwidth]{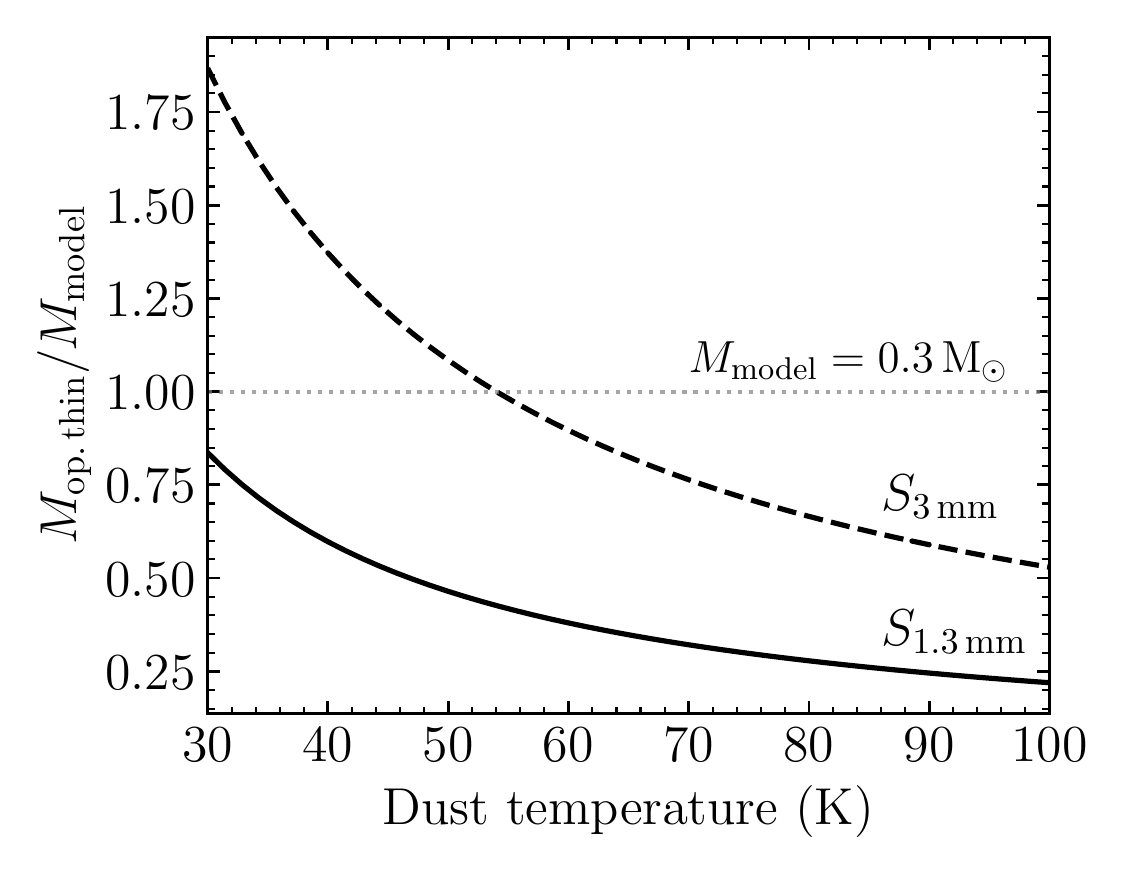}
 \caption{Ratio of the observational mass estimates over the original disk mass, for the RHD model. Masses are derived using equation \ref{eq:disk_mass} from the flux densities at 1.3 and 3\,mm. The figure shows that the mass is best estimated for a dust temperature of 55\,K with the less optically thick emission, i.e., at 3\,mm.}
 \label{fig:observational_mass_estimates}
\end{figure}

The mass of a disk can be estimated from its flux density using the following equation \citep{Hildebrand1983}: 

\begin{equation}
    \label{eq:disk_mass}
    M_{\rm disk} = \frac{g S_{\nu} d^2}{\kappa_{\nu} B(T_{\rm dust})},
\end{equation}

\noindent where $S_{\nu}$ is the flux, $g$ is the gas-to-dust mass ratio of 100, $d$ is the distance, $\kappa_{\rm 1.3\,mm}=1.50\,$cm$^2$\,g$^{-1}$ and  $\kappa_{\rm 3\,mm}=0.58\,$cm$^2$\,g$^{-1}$ are the dust opacities and $B(T_{\rm dust})$ is the Planck function for a given dust temperature $T_{\rm dust}$. 
The fluxes of the RHD model integrated over a 5\,$\sigma$ region are $S_{\rm 1.3\,mm}=1.60\,$Jy and $S_{\rm 3\,mm}=0.29\,$Jy. Figure~\ref{fig:observational_mass_estimates} shows the resultant masses as a fraction of the true mass ($0.3\,{\rm M}_{\odot}$) for temperatures from 30 K to 100 K. 
As the figure shows, although the disk temperatures are high, at least 100 K in the mid-plane (Fig. \ref{fig:dust_temperature_profiles}), when using the optically thin approximation at 1.3mm, the measured mass is closer to the real value for lower temperatures ($\sim$30 K). 
When assuming higher temperatures, the best estimates are obtained when using the 3 mm fluxes. 
For this particular disk model, using the 3 mm fluxes and a dust temperature of 55 K result in a more accurate estimate. 
This analysis shows that one can obtain a reasonable mass estimate under two erroneous assumptions (optically thin and low temperatures) due to the two effects canceling out each other.  
We caution the reader that the selection of this temperature works best for this particular model but it is not necessarily meant to be used for all kind of embedded disks. 


\section{Conclusions}
\label{sec:Conclusions}
In this work, we presented continuum ALMA observations at 1.3 and 3~mm towards the Class 0 protostar IRAS 16293-2422B with resolutions of 88~mas ($\sim$12~au) and 46 mas ($\sim$6.5~au), respectively. 
We analyzed profiles of brightness distribution and provide a spectral index map. 
We compared these with MHD and RHD numerical simulations of a gravitationally unstable disk with masses of 0.1 and 0.3\,M$_{\odot}$, respectively, both formed following the collapse of a dense core. Our results can be summarized as follows:

\begin{itemize}
    \item[1.] The peak brightness temperatures at 1.3 and 3\,mm are 290\,K and 470\,K, respectively. Both peaks appear shifted to the west with respect to the center of the more extended emission which extends up to $r\sim$46 au at 3mm. The spectral index decreases towards the center, starting from $\sim$3 and reaching values less than 2 in the inner $\sim$ 20\,au. 
    
    \item[2.] We performed dust radiative transfer calculations and synthetic observations assuming maximum grain sizes of $a_{\rm max}=1\,\mu$m and 10\,$\mu$m for the two disk models. For these opacities, the albedo of the dust grains is very low and therefore the continuum flux arises mainly from thermal dust emission. The observed fluxes at both wavelengths can be better reproduced with the RHD disk assuming $a_{\rm max}=10\,\mu$m. This model has higher dust temperatures ranging from 200 K to 400 K across the disk midplane. These higher temperatures are naturally achieved by the numerical simulation that include radiative transfer coupled with additional heating from the compression triggered by the spiral arms. 
    
    \item[3.] Radiative heating from the central protostar alone is not able to heat up the dust beyond a few au from the central source and thus, cannot reproduce the high observed brightness temperatures (100 K to 200 K) between $\sim$15-30 au. These results still hold when increasing the luminosity of the central source or changing the dust opacities.
    
    \item [4.] The presence of spiral arms in the RHD model leads to asymmetries in the brightness profiles beyond the peak. These asymmetries resemble the observed ones and thus we speculate spiral the structures could be present in this particular source.
    
    \item[5.] Both disk models reproduce well the low spectral index values and its spatial distribution. This is because both disks are optically thick and show temperatures that are higher in the inner layers compared with the outer layers (i.e., self-obscuration). Thus, low spectral index values in embedded disks can naturally arise due to these disks being optically thick and with an increasing temperature towards the inner layers (including the midplane) unlike the more evolved protoplanetary disks at comparable scales. This scenario does not require mm grain sizes or self-scattering effects.
\end{itemize}

The high temperatures present in the disk around source B allow to explain the large variety of complex organic molecules observed in the hot corino IRAS 16293-2422, as reported by recent spectral lines studies, such as the PILS survey. 
The possibility that these high temperatures are the consequence of gravitational instabilities, points to the idea that self-gravitating disks may provide the conditions for the detection of hot corinos in Class 0 sources. 
Thus, future observational and modelling studies targeting emission of COMs in gravitationally unstable disks may help to test this scenario. 

Future high-resolution continuum observations at different wavelengths are needed to further investigate density and temperature substructures present in the disk. 
Similarly, future multiwavelength high-resolution observations are necessary to investigate if the temperature and density distribution models presented in this work can also explain the fluxes and spectral index behavior for other embedded disks, providing important observational constraints to the physical structure of young protostellar disks.

\section*{Acknowledgements}
J.Z., M.J.M., B.Z. and P.C. acknowledge the support of the Max Planck Society.
H.B.L. is supported by the Ministry of Science and Technology (MoST) of Taiwan (Grant Nos. 108-2112-M-001-002-MY3. J.D.I. acknowledges support from the Science and Technology Facilities Council of the United Kingdom (STFC) under ST/T000287/1.

This paper makes use of ALMA data from the following projects: 2017.1.01247.S (PI: G. Dipierro) and 2016.1.00457.S (PI: Y. Oya). ALMA is a partnership of ESO (representing its member states), NSF (USA) and NINS (Japan), together with NRC (Canada), MOST and ASIAA (Taiwan), and KASI (Republic of Korea), in cooperation with the Republic of Chile.
The Joint ALMA Observatory is operated by ESO, AUI NRAO and NAOJ. 
The National Radio Astronomy Observatory is a facility of the National Science Foundation operated under cooperative agreement by Associated
Universities, Inc.

\section*{Data Availability}
The data underlying this article will be shared on reasonable request to the corresponding author.



\bibliographystyle{mnras}
\bibliography{references} 



\appendix

\section{Spectral index}
\label{appendix:spectral_index}
The slope of the spectral energy distribution (namely, the spectral index $\alpha$) in logspace, between $\nu_1$ and $\nu_2$ (for $\nu_1>\nu_2$), can be obtained as 
\begin{equation}
    \label{eq:spectral_index}
    \alpha = \frac{\log(I_1) - \log(I_2)}{\log(\nu_1) - \log(\nu_2)},
\end{equation}
with $I_{1}$ and $I_{2}$ the specific intensities at frequencies $\nu_1$ and $\nu_2$, respectively. In the Rayleigh-Jeans regime, the relation between the intensity and the brightness temperature $T_{\rm b}$ at frequency $\nu$ is given by  
\begin{equation}
    \label{eq:flux-Tb_relation}
    I_{\nu} = \frac{2 k_{\rm B}}{c^2} \nu^2 \, T_{\rm b},
\end{equation}
with $k_{\rm B}$ the Boltzmann constant. Since $I\propto \nu^2T_{\rm b}$, then equation \ref{eq:spectral_index} can be expressed as
\begin{eqnarray}
    \alpha &=& \frac{\log\left(\nu_1^2 \, T_{\rm b, 1}\right) - \log\left(\nu_2^2 \, T_{\rm b, 2}\right)}{\log\left(\nu_1\right) - \log\left(\nu_2\right)}, \\
           &=& \frac{2\log\left(\frac{\nu_1}{\nu_2}\right)}{\log\left(\frac{\nu_1}{\nu_2}\right)} + \frac{\log\left(\frac{T_{\rm b, 1}}{T_{\rm b, 2}}\right)}{\log\left(\frac{\nu_1}{\nu_2}\right)}, \\
           &=& 2 + \frac{\log(T_{\rm b, 1}) - \log(T_{\rm b, 2})}{\log(\nu_1) - \log(\nu_2)}, 
\end{eqnarray}
thus,
    
    \[ \centering T_{\rm b, 1} 
        \begin{cases} 
          \,\, \geqslant T_{\rm b, 2} & , \,\, \alpha \geqslant 2\\
          \,\, < T_{\rm b, 2} & , \,\, \alpha < 2\,\, .
       \end{cases}
    \]
This last inequality tells us that a spectral index observed to be lower than 2 means that the brightness temperature at higher frequency is lower than at lower frequency, which is the opposite to what is expected for a black-body radiator at a single temperature. 
In our observations this low value of $\alpha$ can be explained by the two wavelengths tracing optically thick ($T_{\rm b}\lesssim T_{\rm BB}$) layers at different temperatures. 
Moreover, since the opacity in the Rayleigh-Jeans regime decreases as a function of frequency ($\kappa\propto\nu^{\alpha-2}$), then the surface at $\tau=1$ traced by the 3~mm observation lies deeper into the disk as compared to that traced at 1.3~mm. 
This is indicative of the inner regions of the disk being warmer than the outer regions.

\section{Spectral index for the MHD disk model}
\label{appendix:spectral_index_bo_model}
\begin{figure*}
    \centering
    \includegraphics[width=18cm, trim=1cm 0 1cm 0cm, clip]{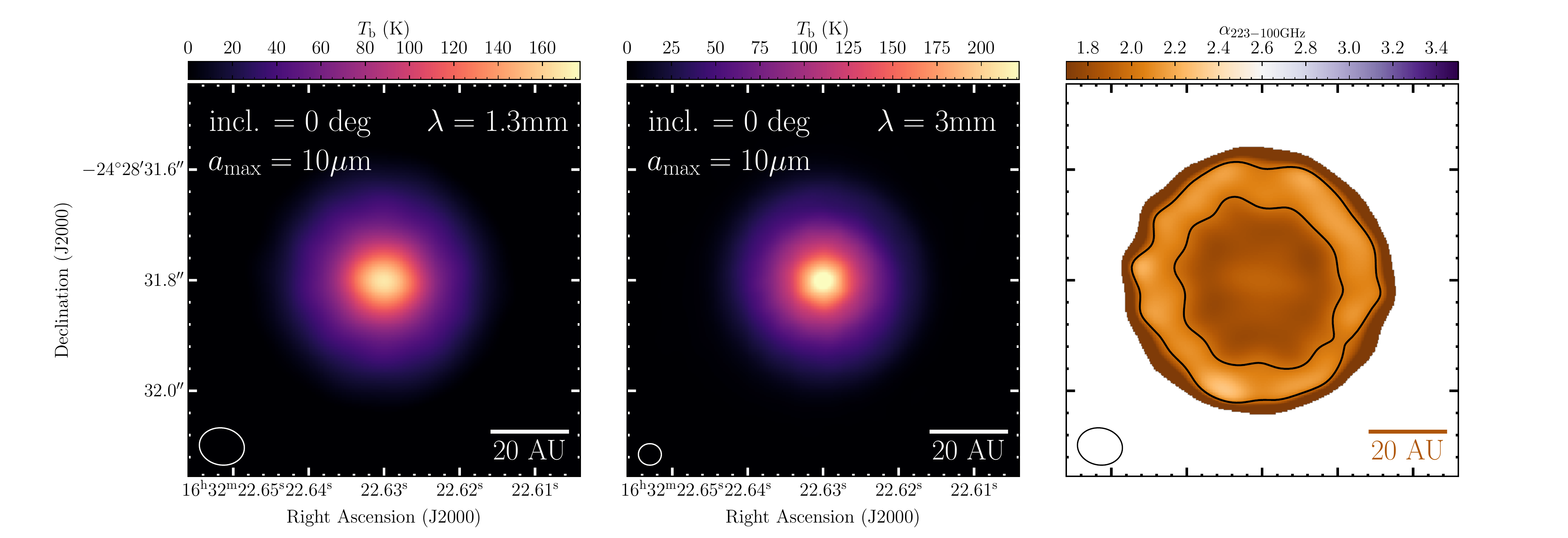}
    \caption{Synthetic brightness temperature maps at 1.3 and 3~mm along with the spectral index for the MHD disk formation model from~\citep{Zhao2018}, described in section~\ref{sec:non-ideal_MHD_model}. A similar behaviour is observed as compared to the RHD gravitationally unstable model, in which the spectral index ($\alpha$) is observed to decrease toward the center of the disk. The contours in the rightmost panel indicate $\alpha=1.7$ and 2, inside-out.}
    \label{fig:spetral_index_bo_model}
\end{figure*}
To complement the analysis presented in section \ref{sec:simulated_observations_grav-unstable_model}, we have also generated the 1.3~mm, 3~mm and spectral index maps for the MHD disk formation model described in section~\ref{sec:non-ideal_MHD_model}. 
The results are shown in Fig.~\ref{fig:spetral_index_bo_model} in the same layout as Figs.~\ref{fig:IRAS16293B_observations} and \ref{fig:simulated_observations}. 
We observed a similar tendency as compared to the RHD model in which the spectral index falls to around 1.8 toward the densest parts of the model.
This implies that both models considered in this study, are capable of reproducing the low values of $\alpha$ expected from early disks with a midplane at higher temperatures than the outer layers.

\section{Toomre Q parameter for both disk models}
\label{appendix:toomre_parameter_for_both_models}

\begin{figure}
 \centering
 \includegraphics[width=1.0\columnwidth]{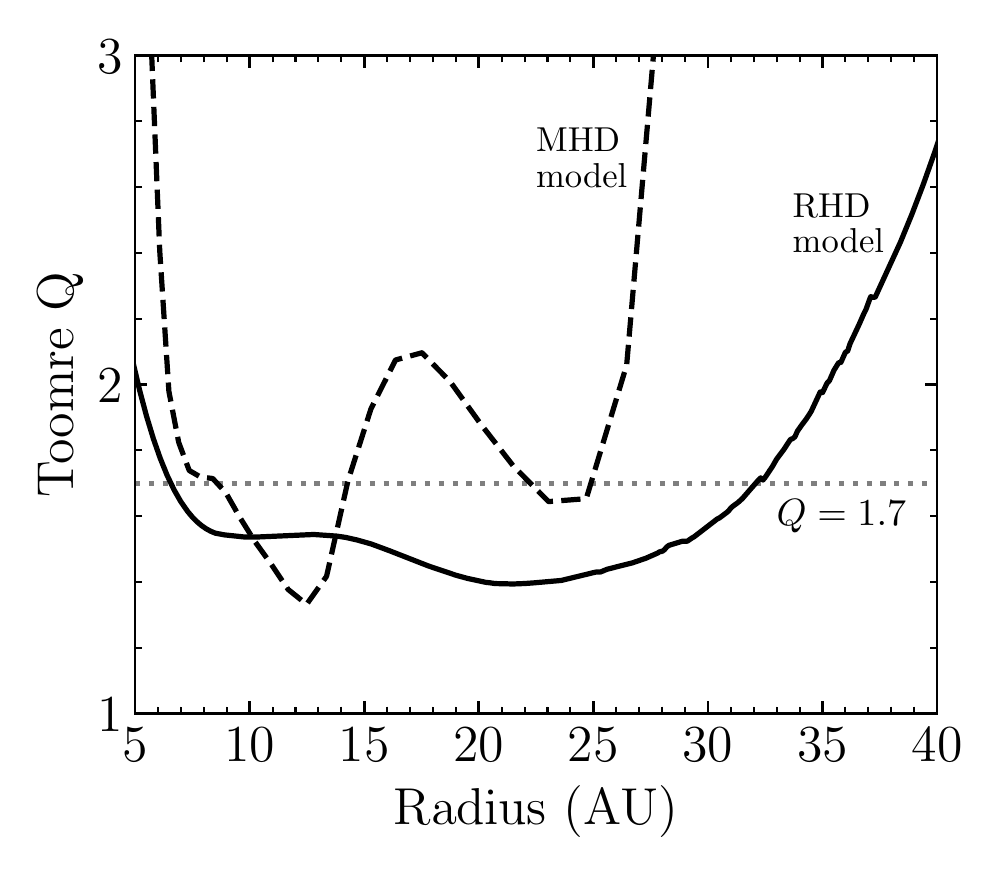}
 \caption{Toomre Q parameter of both disk models. The RHD model has a Q$\lesssim1.7$ at radii larger than 7\,au, with a minimum value of 1.4 at 20\,au. The MHD disk also reach values lower than 1.7 between 8 and 25\,au. In spite of the spiral arms being prominent in only one of the two models (RHD), both disk are actually gravitationally unstable.}
 \label{fig:toomre_q}
\end{figure}

In Fig. \ref{fig:toomre_q} we show the Toomre parameter for the two disk models considered here, the RHD and MHD disk. Both disks are gravitationally unstable and susceptible to non-axisymmetric perturbations in the regions where Q$\lesssim$1.7 \citep{Durisen2007}, which is between 7 and 20\,au for the RHD disk and between 8 and 25\,au for the MHD, with a small exception within a radius of 15 to 22\,au, but still remaining under Q=2.  In spite of the spiral arms being prominent in only one of the two models (RHD), both disk are actually gravitationally unstable.


\bsp	
\label{lastpage}
\end{document}